\pdfoutput=1
\documentclass[12pt]{article}
\usepackage[margin=1in]{geometry}
\usepackage{amsmath}
\usepackage{titling}
\usepackage{blindtext}
\usepackage{pgfplots}
\usepackage{natbib}
\def\sin{\mathop{\rm sin}\nolimits}
\usepackage{comment}
\usepackage{setspace}
\usepackage{authblk}
\usepackage[inline]{trackchanges} 
\usepackage[percent]{overpic}
\usepackage[section]{placeins}
\usepackage{pbox}
\usepackage{comment}
\usepackage{titling}
\usepackage{natbib}
\usepackage{soul}
\usepackage{siunitx}
\usepackage{textcomp}
\newcommand{\GG}[1]{}

\usepackage{ragged2e}
\justifying

\begin{document}

\title{Application of a Modified Spheromak Model to Simulations of Coronal Mass Ejection in the Inner Heliosphere}

\author[1]{Talwinder Singh}
\affil[1]{Department of Space Science, The University of Alabama in Huntsville, AL 35805, USA}

\author[2]{Tae K. Kim}
\affil[2]{Center for Space Plasma and Aeronomic Research, The University of Alabama in Huntsville, AL 35805, USA}

\author[1,2]{Nikolai V. Pogorelov}

\author[3]{Charles N. Arge}
\affil{Solar Physics Lab, NASA/GSFC, Greenbelt, MD 20771, USA}

\setcounter{Maxaffil}{0}
\renewcommand\Affilfont{\itshape\small}
\date{}  

\begin{titlingpage}
    \maketitle

\begin{abstract}
 {The magnetic fields of interplanetary coronal mass ejections (ICMEs), which originate close to the Sun in the form of a flux rope, determine their geoeffectiveness.} Therefore, robust flux rope-based models of CMEs are required to perform magnetohydrodynamic (MHD) simulations aimed at space weather predictions. We propose a modified spheromak model and demonstrate its applicability to CME simulations. In this model, such properties of a simulated CME as the poloidal and toroidal magnetic fluxes, and the helicity sign can be controlled with a set of input parameters. We propose a robust technique for introducing CMEs with an appropriate speed into a background, MHD solution describing the solar wind in the inner heliosphere.  Through a parametric study, we find that the speed of a CME is much more dependent on its poloidal flux than on the toroidal flux. We also show that the CME speed increases with its total energy, giving us control over its initial speed. We further demonstrate the applicability of this model to simulations of CME-CME collisions. Finally, we use this model to simulate the 12 July 2012 CME  and compare the plasma properties at 1 AU with observations. The predicted CME properties agree reasonably with observational data. 
\end{abstract}
\end{titlingpage}

\section{Introduction}\label{introduction}
Coronal Mass Ejections (CMEs) are one of the most explosive events in our solar system, with the kinetic energy release of up to  {$10^{26}$} J \citep{Forbes00,Vourlidas02}. During a CME event, the coronal plasma erupts producing a coherent magnetic structure called the flux rope. The standard model of CME eruption states that  {the pre-eruptive state possesses a core of axial flux that may be either a sheared arcade or a }flux rope which exists above the magnetic polarity inversion line \citep[][]{Carmichael1964, Sturrock1968, Hirayama1974, Kopp1976}. As this flux rope starts to rise due to various possible magnetic force imbalances, the overlying magnetic field lines serve as the dominant source of the poloidal flux in a flux rope \citep[see the review of][]{Chen11}.  {As a result, a CME carrying mass between $10^{14}$ to $4\times10^{16}$ gms, starts its journey through the heliosphere with speeds ranging between 10 to $>$2000 km/s \citep{Hudson2006}}. Most of the time, the CME speed exceeds the fast magnetosonic wave speed in the ambient solar wind, resulting in a shock in front of this CME \citep{Sime87, Raymond00}. 

 {If a CME impacts Earth, it disturbs the local space environment posing a wide range of risks including harmful effects on space assets, radiation exposure for astronauts and passengers during polar flights, communication losses between satellites and ground receivers, and inducing large currents in long power transmission lines.} The ability of a CME to disturb the near-Earth space, also called its geo-effectiveness, is largely due to CME's magnetic  {field which often reveals itself as a flux rope with a sheath region in front of it. CMEs observed to have flux ropes associated with magnetic clouds at 1 AU form a subset of the total CME variety \citep{Manchester2017}.} If the orientation of the magnetic field in the CME is favorable for magnetic reconnection with the Earth's magnetosphere, such a CME is highly likely to be geoeffective \citep{Burton75, Gonzalez94}. This orientation should involve a negative $B_z$ component, since $B_z$ is positive in the Earth's day-side magnetosphere. 

To mitigate problems caused by geoeffective CMEs with proper precautionary steps, we need to predict accurately the arrival time and magnetic structure of CMEs at 1 AU. Due to the development of supercomputers and highly parallelized codes, magnetohydrodynamic (MHD) simulations of CMEs have become a major tool for CME predictions. A number of authors have shown the application of MHD modeling to CME simulations  {\citep[e.g.][]{Vandas1996, Vandas1997, Manchester04, Lugaz05, Aulanier10, Shen14, Titov14, Jiang16, Shiota16, Pogorelov17, Jin17, Singh18, Scolini19, An2019}}. The cone model of a CME \citep[e.g.][]{Chane05,Odstrcil99} is already being used for operations by various agencies.  {This model can predict the CME arrival time, shock and sheath regions while its magnetic structure remains unknown because of the lack of a flux rope treatment.} This major disadvantage of the cone model can be addressed by using flux-rope-based models. Many flux rope models have been proposed so far for CME simulations, e.g., 1) the spheromak model \citep{Lites95}, 2) the Gibson-Low model \citep{Gibson98}, 3) the Titov--Demoulin model \citep{Titov99}, etc. 

For a successful prediction of CMEs, the input parameters in a flux rope model must be constrained  by observations as much as possible. Keeping that in mind, \cite{Singh20} proposed a modified spheromak model based on the observed poloidal flux, toroidal flux, and helicity sign. The parameters of this flux rope can be adjusted so that it erupts with the observed speed, orientation, and direction. In this paper, we show the applicability of this model to simulate CMEs in the inner heliosphere, where the inner boundary of the computation domain is at 0.1 AU.  {It is important that model CMEs should propagate through a realistic data-driven background solar wind (SW) because the CME-SW interaction plays an important role during the CME propagation.} We use the inner-heliospheric background created on the basis of the Wang--Sheeley--Arge (WSA) coronal model as inner boundary conditions \citep{Arge00, Arge03, Arge04, Arge05}. 

Here, we perform a parametric study to analyze how a CME possessing the modified spheromak flux-rope structure evolves through the inner heliosphere. We show the effect of varying poloidal and toroidal fluxes, initial flux rope size, and total energy on the CME dynamics. We also show the applicability of our model to simulations of CME-CME interactions, which are very common and important during the solar activity maxima. We also show the simulation results for the 12 July 2012 CME and compare simulation results with observations at 1 AU.

In Sec. \ref{models}, we describe the SW and flux rope models used in this study. We also describe how the flux rope model is inserted into the solar wind background. In Sec. \ref{results}, we discuss the results of various CME runs performed in this study. In particular, the CME-CME collision results are presented in this section. We also show the results for the 12 July 2012 CME simulation. Finally, we give our concluding remarks in Sec. \ref{conclusion}.

\section{Models}\label{models}
  {All our simulations are carried out using the Multi-Scale Fluid-Kinetic Simulation Suite (MS-FLUKSS), a highly paralleled collection of codes capable of performing AMR finite-volume MHD simulations of the SW in the presence of neutral atoms and physical processes,e.g. turbulence, induced by charge exchange on adaptive mesh refinement grids \citep{Pogorelov14}.} In the following subsections, we describe our SW and flux-rope CME models, and the method of CME insertion into the SW background.

\subsection{Inner Heliosphere Model}

To generate realistic time-varying background solar wind in the inner heliosphere, we couple our heliospheric MHD model with the WSA coronal model at 0.1 AU \citep[e.g.,][]{Kim19}. As input to the WSA model at 1 $R_\odot$, we use the Air Force Data Assimilative Photospheric Flux Transport (ADAPT) synchronic maps generated from the NSO/GONG magnetograms \citep{Arge10,Arge11,Arge13,Hickmann15}. The semi-empirical WSA model initially extrapolates the photospheric magnetic field to a source surface, chosen to be at 2.5 $R_\odot$, using the potential field source surface (PFSS) model, and then to the outer boundary at 21.5 $R_\odot$ (0.1 AU) using the Schatten current sheet model. In addition to magnetic field strengths, the WSA model provides an estimate of the solar wind speed at the model outer boundary as a function of the flux expansion factor and distance to the nearest coronal hole boundary \citep{Arge03,Arge05}.

\begin{figure}[!htb]
\center
\includegraphics[scale=0.1,angle=0,width=15cm,keepaspectratio]{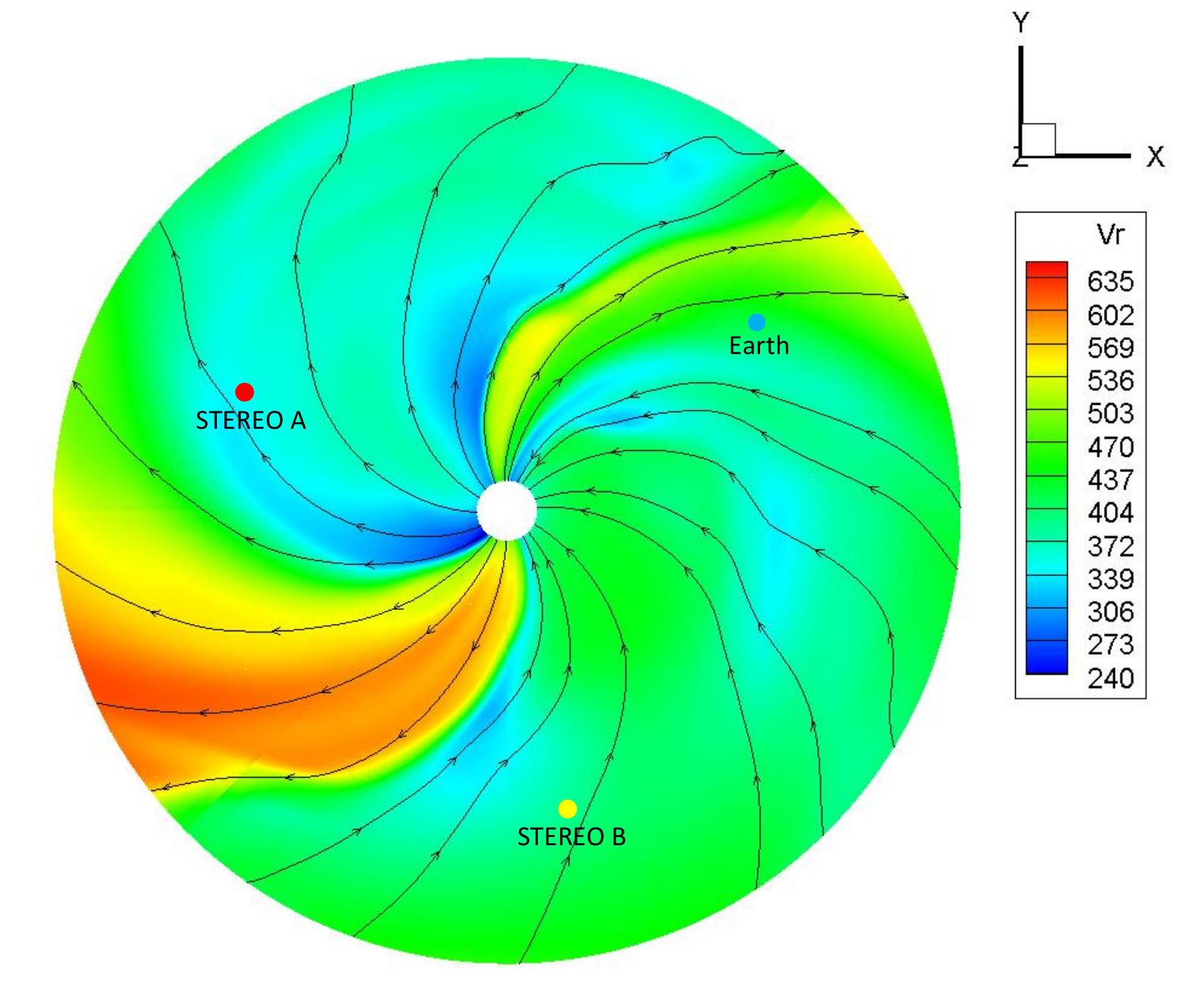}
\caption{Radial velocity of the solar wind and magnetic field lines in the inner heliosphere on 12 July 2012 21:30 UT is shown in the equatorial plane between 0.1 and 1.5 AU.}
\label{IH_back}
\end{figure}

While the ADAPT-WSA model provides 12 realizations of magnetic field and solar wind speed maps at 0.1 AU, we select one particular realization that provides us with the best fit to the near-Earth spacecraft data.
On interpolating the WSA maps from the original $144 \times 72$ ($2.5^\circ \times 2.5^\circ$) resolution to a base grid of $256 \times 128$ ($\sim 1.4^\circ \times 1.4^\circ$) in MS-FLUKSS, we scale the WSA magnetic field strengths by a factor of 3 to compensate for the systematic underestimation of the open magnetic flux at 1 AU \citep{Linker16,Linker17,Wallace19}. We estimate the radial and azimuthal components of magnetic field at 21.5 $r_\odot$ using the local solar wind speed to account for the solar rotation as the solar wind propagates radially outward from 1 to 21.5 $r_\odot$ \citep{MacNeice11}. Also, we reduce the WSA speeds by 20\% to account for the difference in solar wind acceleration between the WSA and MS-FLUKSS models \citep[e.g.,][]{MacNeice11,Kim14}. To estimate the solar wind density (and temperature) at 0.1 AU, we use empirical correlations between the solar wind speed and density (and temperature) based on OMNI data \citep{Elliott16}. With these boundary conditions, we solve the ideal MHD equations on the nonuniform $150 \times 256 \times 128$ ($r$, $\phi$, $\theta$)  {fully} spherical grid to simulate the solar wind outflow from 0.1 to 1.5 AU. The resulting solar wind solution in the equatorial plane is shown in Fig. \ref{IH_back}. Here, we have colored the equatorial plane with radial solar wind speed. We have also shown the magnetic field lines in this plane. 

\subsection{Modified spheromak model}
\citet{Singh20} describe the details of the modified spheromak model used in this study. The original spheromak solution was modified so that the poloidal and toroidal fluxes can be set up independently. The analytic solution for the magnetic field inside the flux rope can be given as
\begin{equation}\label{Eq1}
\vec{b} = \frac{1}{r \sin \theta}\Big(\gamma \frac{1}{r}\frac{\partial A}{\partial \theta}\hat{r} - \gamma \frac{\partial A}{\partial r}\hat{\theta} + \delta \alpha_0 A \hat{\phi}\Big),     
\end{equation}
\begin{equation}
A = \frac{4 \pi a_1}{\alpha_0^2}\Big[\frac{r_0^2}{g(\alpha_0 r_0)}g(\alpha_0 r) - r^2 \Big] \sin^2 \theta , 
\end{equation}
\begin{equation}
g(\alpha_0 r) = \frac{\sin (\alpha_0 r)}{\alpha_0 r} - \cos (\alpha_0 r) , 
\end{equation}
where $\alpha_0$ and $r_0$ are related as $\alpha_0 r_0 = 5.763459$, which is the first root of the Bessel function $J_{5/2}$. Here, $r_0$ is the spheromak radius. The toroidal flux in the spheromak is proportional to the parameter $a_1$, whereas the poloidal flux is proportional to $a_1 \gamma$.  The origin of the spherical coordinate system $(r,\, \theta,\, \phi)$ is placed at the spheromak center. The parameter $\delta$ can assume values +1 or -1 for either positive and negative helicity, respectively. The plasma density is distributed uniformly in the spheromak.  {The thermal pressure is assumed to be proportional to the magnetic pressure through the choice of desired plasma $\beta$. Our approach to do this will be discussed in the next subsection}. As suggested by \citet{GL98}, this model can also undergo a stretching operation $r\rightarrow r-a$, where $a$ is the stretching parameter to modify the spheromak shape from spherical to a tear-drop. This stretching operation does not violate the solenoidal condition for magnetic field. 

\subsection{Introducing flux rope in solar wind}\label{FRinSW}
We propose to introduce a flux rope into the SW in such a way that initially the flux rope is superimposed with the background. This approach is different from the one commonly used, where a flux rope is introduced gradually at the inner boundary \citep[eg.][]{Shiota16, Scolini19}. We are not introducing the full spheromak into the SW, but rather the top half of it, which resembles the CME flux rope much more accurately, exhibiting a curved front and two legs. We introduce the flux rope parameters as follows:
\begin{itemize}
    \item $\vec{b}_{final} = \vec{b}_{FR}$,
    \item $\rho_{final} = \rho_{FR} + \rho_{SW}$,
    \item $e_{final} = \xi e_{FR} + e_{SW}$.
\end{itemize}

Here, $\vec{b}_{FR}$ is given by Eq. (\ref{Eq1}), $\rho$ is the plasma density, and $e$ is the total energy density. We also note that
\[
e_{FR} = \frac{|\vec{b}_{FR}|^2}{8\pi}, \quad e_{SW} = \frac{p_{SW}}{\gamma - 1} + \frac{|\vec{b}_{SW}|^2}{8\pi} + \frac{\rho_{SW} |\vec{v}_{SW}|^2}{2},
\]
where $p$ and $\vec{v}$ are the thermal pressure and bulk velocity, respectively.  {We have kept the isotropic index $\gamma=1.5$ in this study. We also introduce an energy multiplier factor $\xi$ here, which can be shown to be related to plasma $\beta$ as $\beta\approx(\gamma-1)(\xi-1)$ under a quite realistic assumption that the magnetic pressure created by the flux rope is much greater than the ambient solar wind pressure. Thus we can control the plasma $\beta$, as well as the eruption speed of the flux rope, using $\xi$, as will be shown later.} During this insertion procedure, we make sure that total thermal pressure in the domain does not fall below $25\%$ of the background SW thermal pressure \citep{Manchester04}.

Introducing flux rope by modifying the magnetic field in the domain will break the divergence free condition at the flux rope-solar wind interface. This is true even when introducing the flux rope gradually at the inner boundary \citep{Shiota16}. We then rely on divergence cleaning mechanisms like 8-wave approach \citep{Powell99}, generalized lagrange multiplier approach \citep{Dedner02}, etc. to convect away and damp the non-zero magnetic divergence. 

\begin{figure}[!htb]
\center
\begin{tabular}{c c}  
\includegraphics[scale=0.1,angle=0,width=5cm,keepaspectratio]{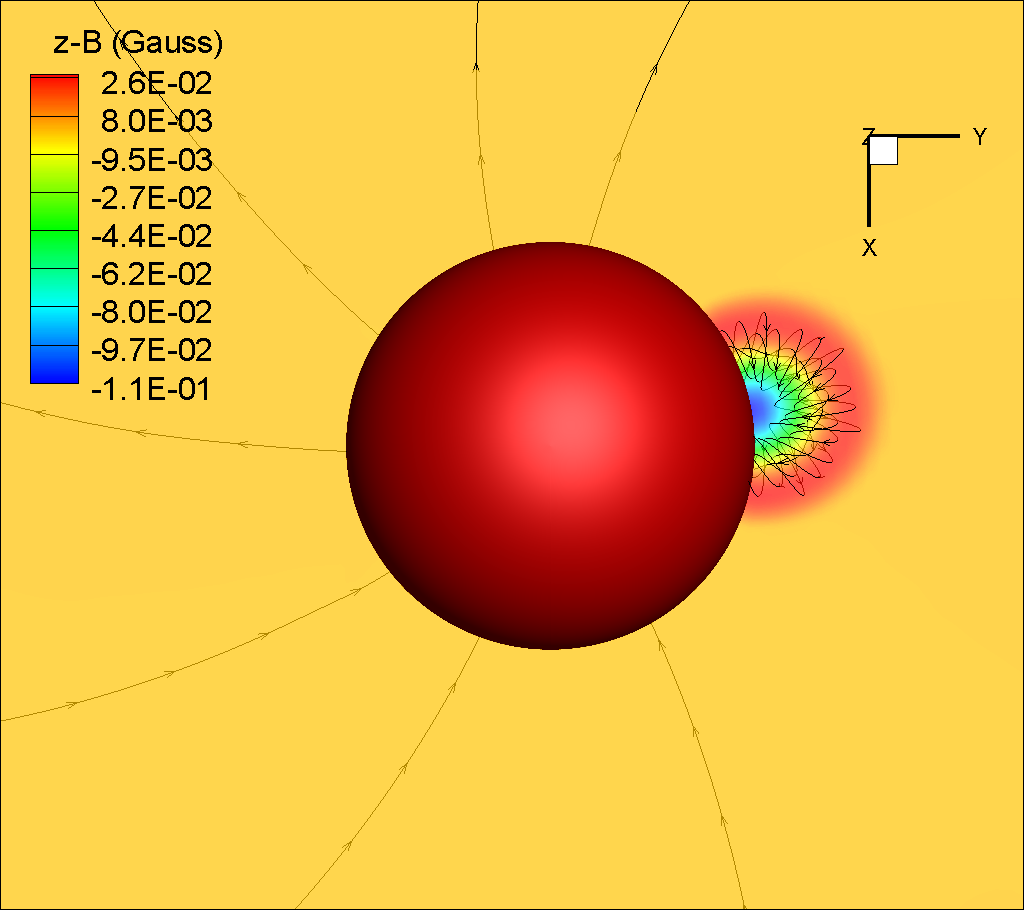}
\includegraphics[scale=0.1,angle=0,width=5cm,keepaspectratio]{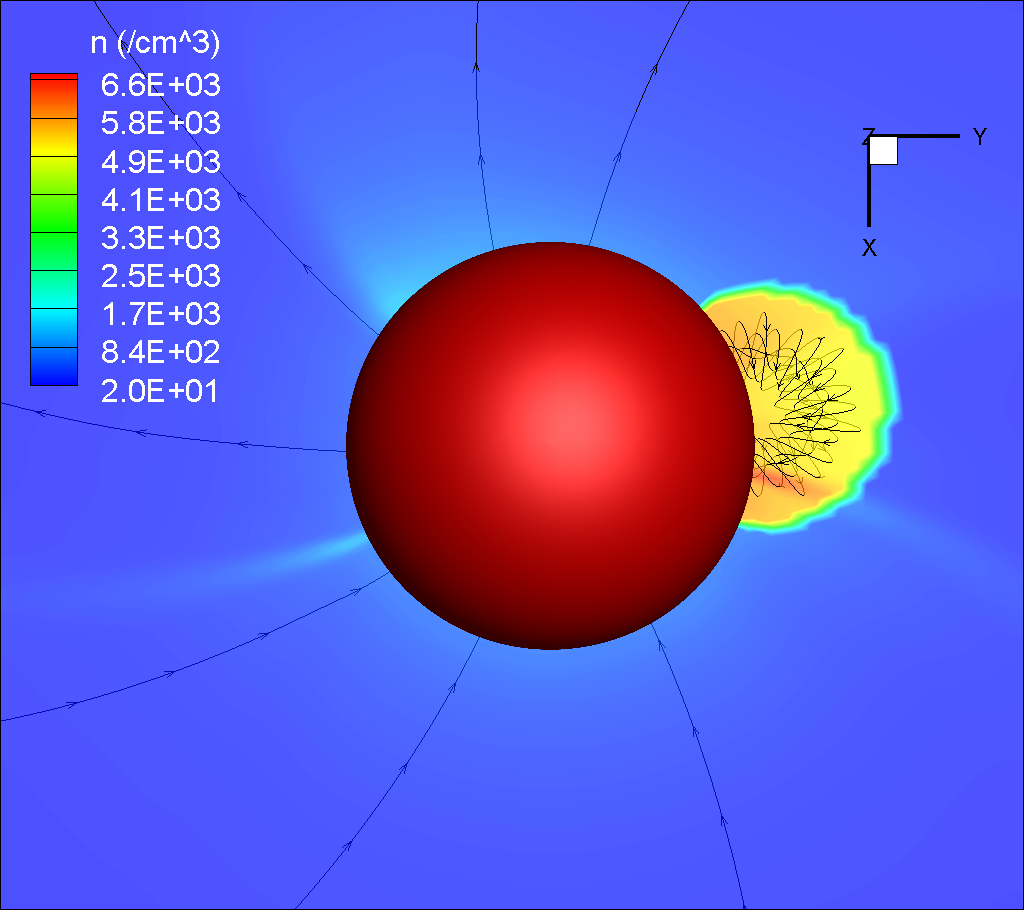}\\
\includegraphics[scale=0.1,angle=0,width=5cm,keepaspectratio]{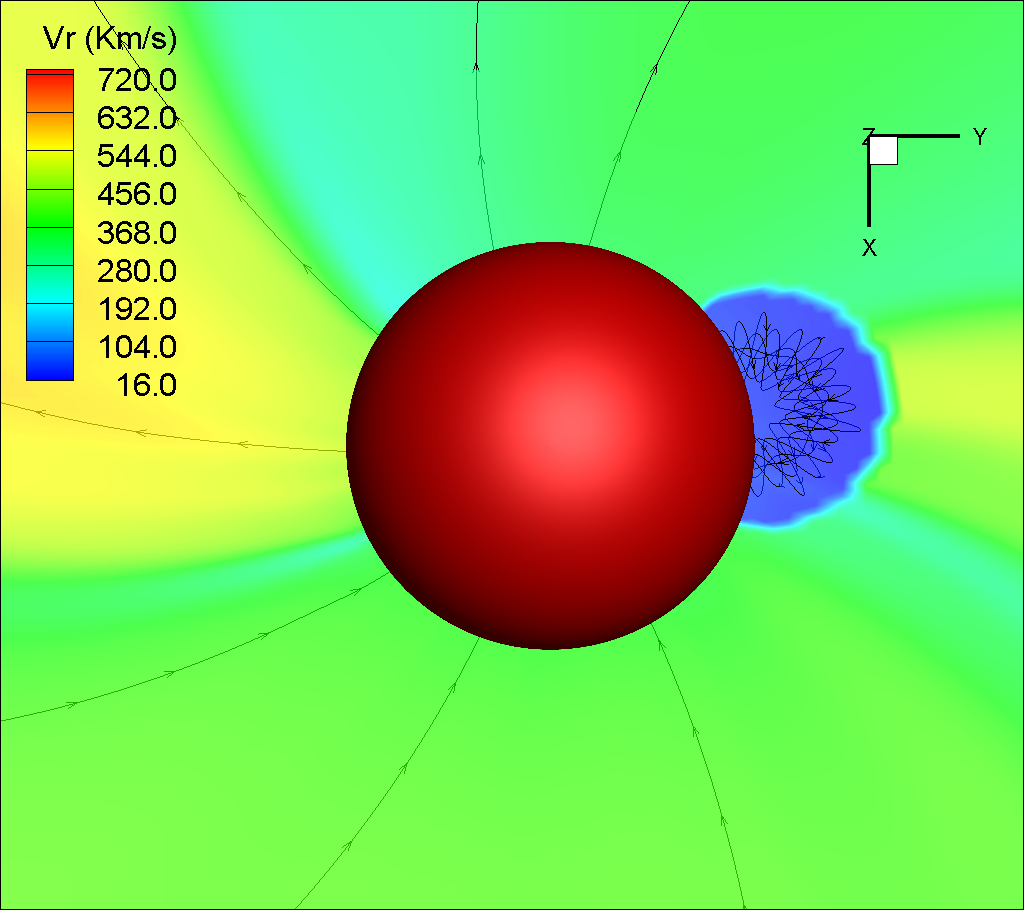}
\includegraphics[scale=0.1,angle=0,width=5cm,keepaspectratio]{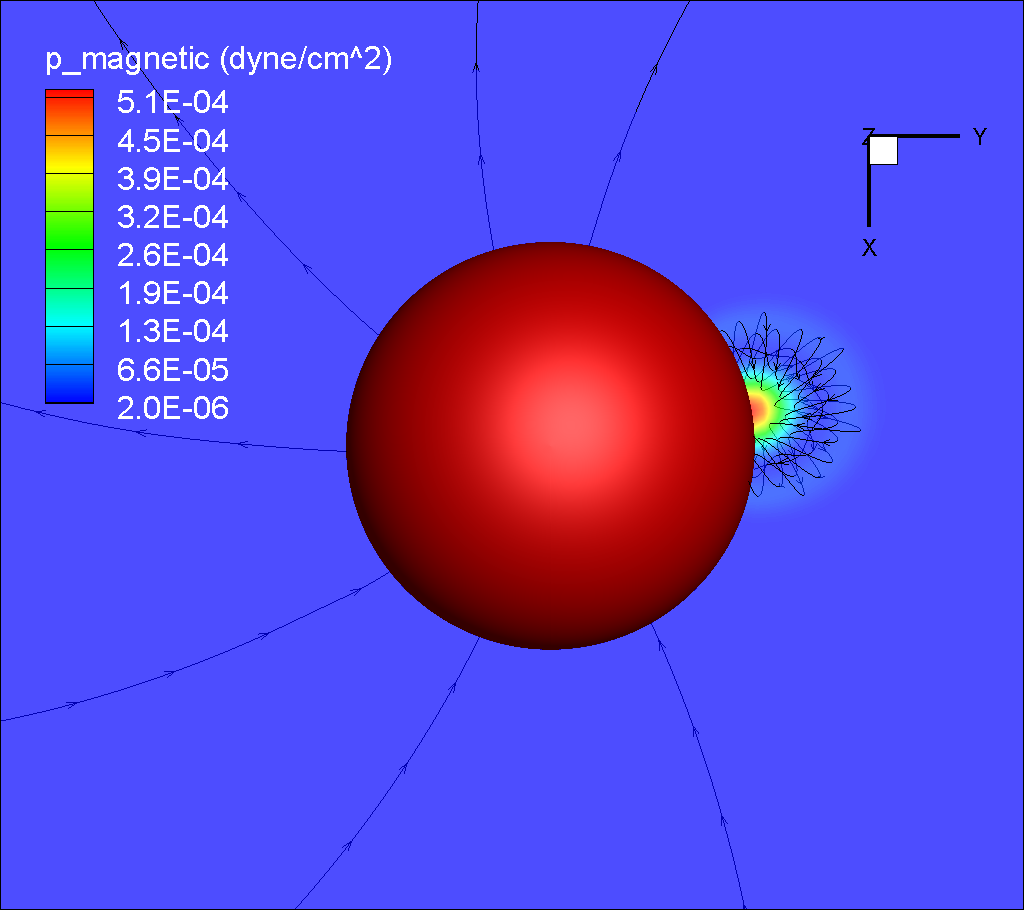}\\
\includegraphics[scale=0.1,angle=0,width=5cm,keepaspectratio]{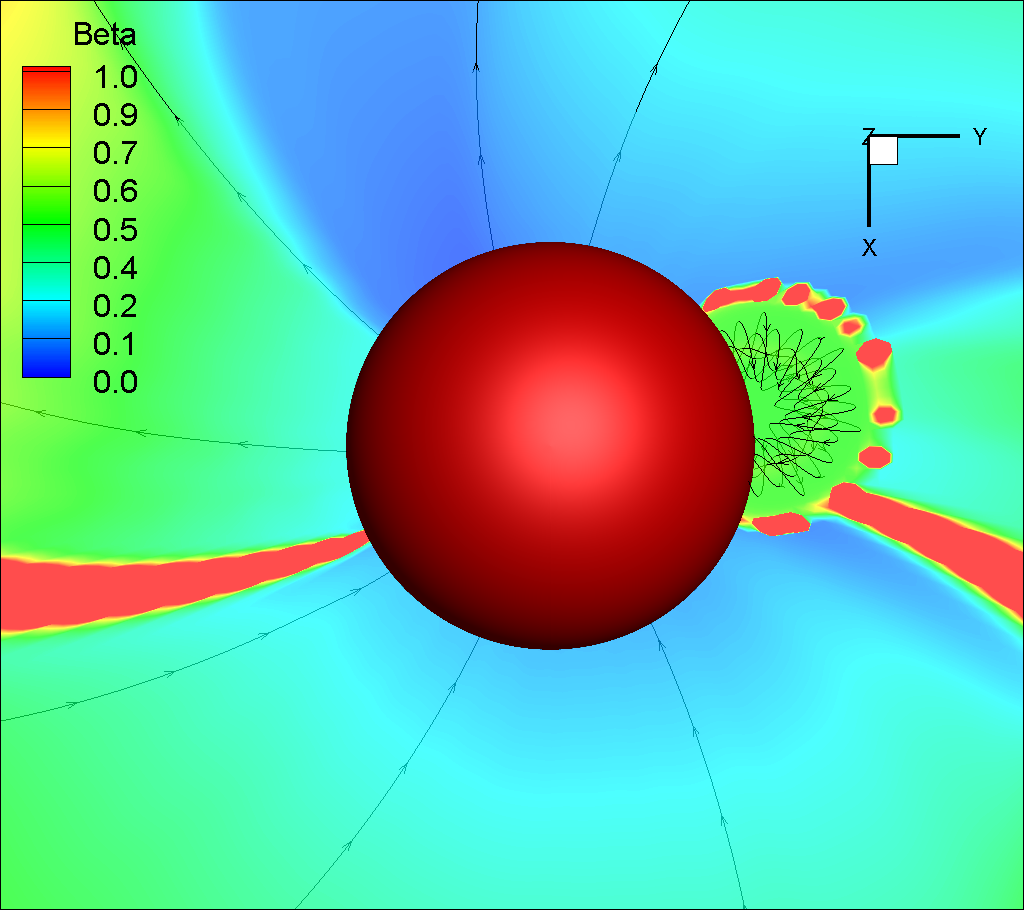}
\includegraphics[scale=0.1,angle=0,width=5cm,keepaspectratio]{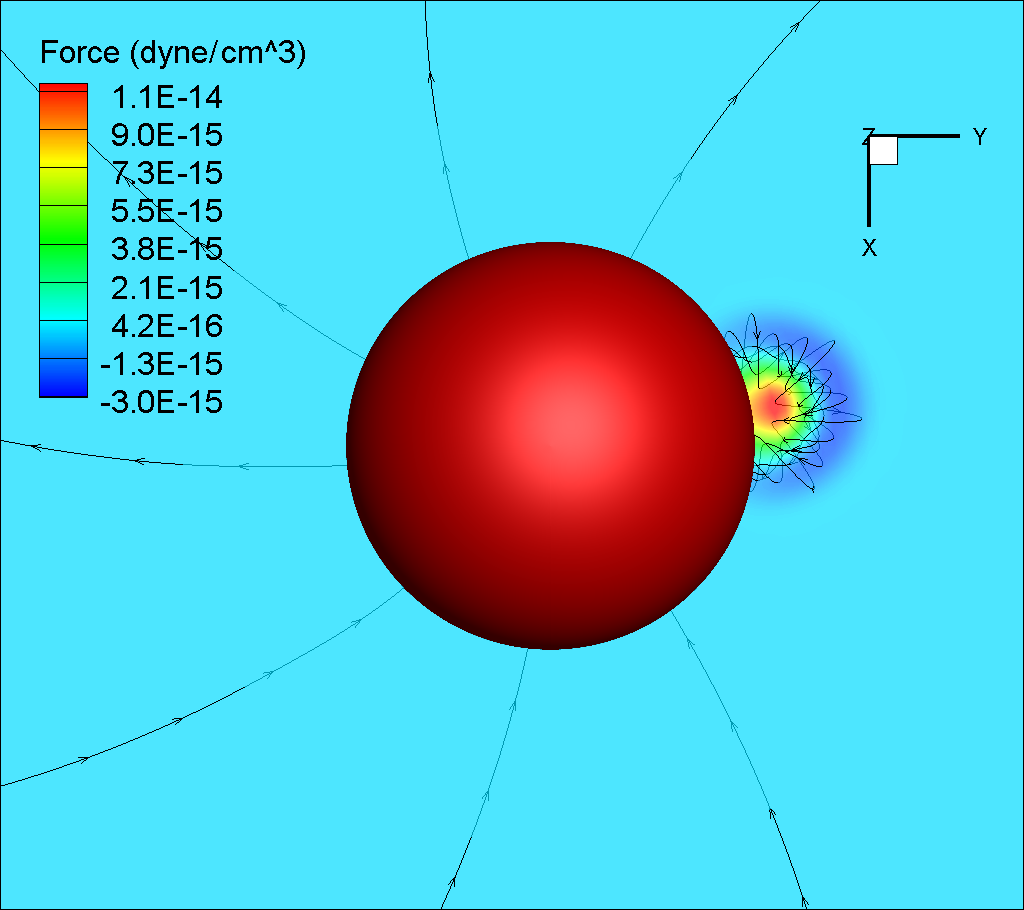}\\
\end{tabular}
\caption{ {Magnetic field line configuration of the modified spheromak inserted at the inner boundary $R=0.1$~AU, shown with the red sphere. We also show various translucent slices through the spheromak showing following parameters:
(top left) $B_z$ in $Gauss$, (top right) ion density in $cm^{-3}$, (middle left) radial speed in $km/s$, (middle right) magnetic pressure in $Dyne/cm^2$, (bottom left) plasma $\beta$, and (bottom right) radial force per unit volume.}}
\label{FR_model}
\end{figure}

The result of the spheromak insertion into the SW is shown in Fig. \ref{FR_model}. The flux rope is kept in the equatorial plane without any tilt. The introduced poloidal flux is $2\times10^{22}$ Mx and the toroidal flux is $5\times 10^{21}$ Mx. The helicity sign is set to be positive and the mass of $1.65\times 10^{15}$ g is distributed uniformly inside it. The red sphere shows the inner boundary placed at 21.5 $r_\odot$. The flux rope size parameter $r_0$ and stretching parameter $a$ are set to be 15 $r_\odot$ and  5 $r_\odot$, respectively.  {The energy multiplier $\xi$ has been set to 2.} The flux rope center is kept at the inner boundary. We find that keeping the flux rope center near the inner boundary creates a configuration suitable for eruption.  {There is a region of high total magnetic pressure inside the flux rope, which can result in its eruption. We notice that for $\xi=2$, plasma $\beta$ is around 0.5, which satisfies our relation $\beta\approx(\gamma-1)(\xi-1)$. This value of $\gamma$ means that magnetic pressure will be the dominant driving force during CME eruption. To visualize the force distribution in this flux rope, we calculated the total force (per unit volume) due to the magnetic pressure gradient, thermal pressure gradient and magnetic tension as shown below:
$$\vec{F}=\frac{\vec{J}\times\vec{B}}{c}-\nabla p,$$
where $$\vec{J}=\frac{c}{4\pi}\nabla\times\vec{B}.$$ 
On solving these equations, we get 
\begin{equation}\label{force_eqn}
\vec{F}=-\nabla\left(\frac{B^2}{8\pi}\right)+\frac{(\vec{B}\cdot \nabla) \vec{B}}{4\pi} - \nabla p
\end{equation}
The first term on the RHS is due to the magnetic  pressure gradient, similarly to the hoop force described by \citet{Kliem2006} for torus instability. The middle term is due to the magnetic tension force, whereas the last term is due to thermal pressure gradient. The radial component of force $\vec{F}$ is shown in the bottom right panel of Fig. \ref{FR_model}. One can see a large radial force underneath the FR, that makes our model CME to erupt. We will show the erupting stages in the next section.}   

\section{Results}\label{results}

  {In this study, we solve the ideal MHD equations in an inertial coordinate system centered at the Sun, whose sidereal rotation period is 25.38 days.} We use a TVD, finite-volume Roe scheme with arithmetic averaging to compute the numerical fluxes and the forward Euler scheme for time integration. To remove numerically-induced magnetic field divergence, we use the \citet{Powell99} approach. We use the WSA solution for the period between 01-June-2012 and 31-Aug-2012 as  {rotating} boundary conditions at the inner sphere to create the time-dependent inner-heliospheric background. We use the spherical domain of size $150\times 128 \times 256$ in $r$, $\theta$, and $\phi$ directions, respectively. 

\begin{figure}[!htb]
\center
\begin{tabular}{c c}  
\includegraphics[scale=0.1,angle=0,width=6cm,keepaspectratio]{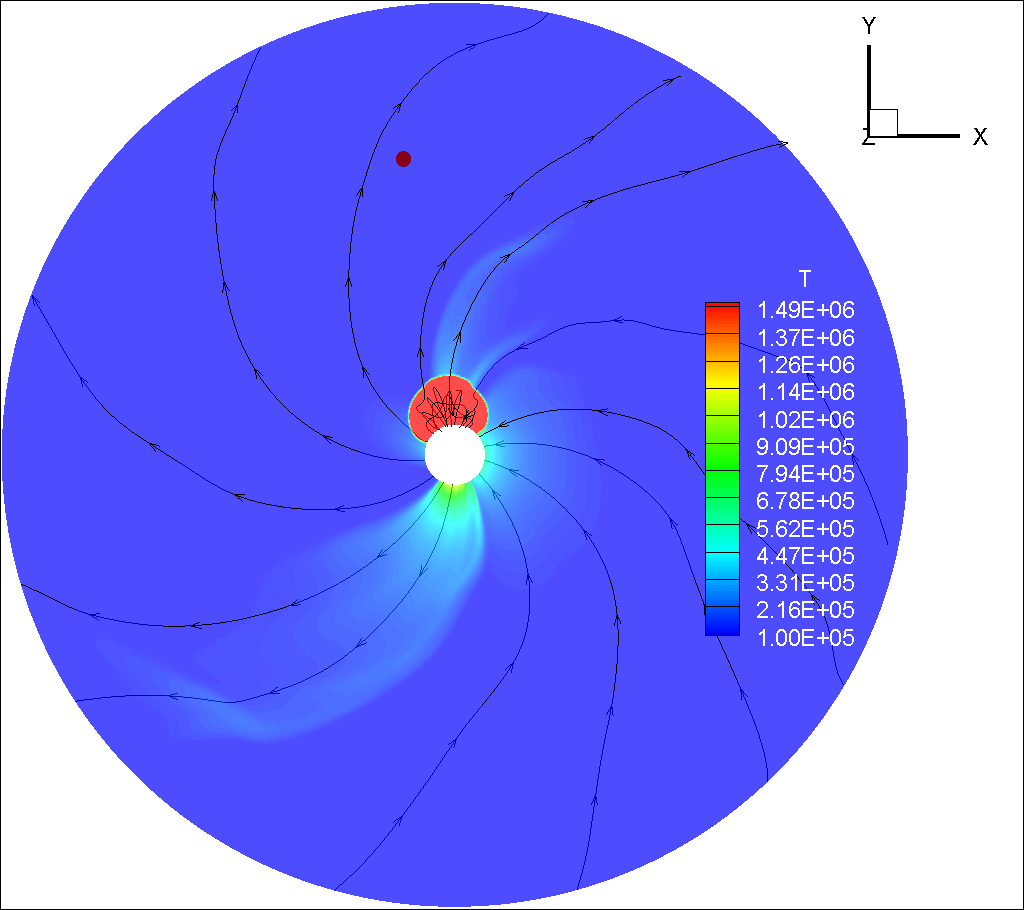}
\includegraphics[scale=0.1,angle=0,width=6cm,keepaspectratio]{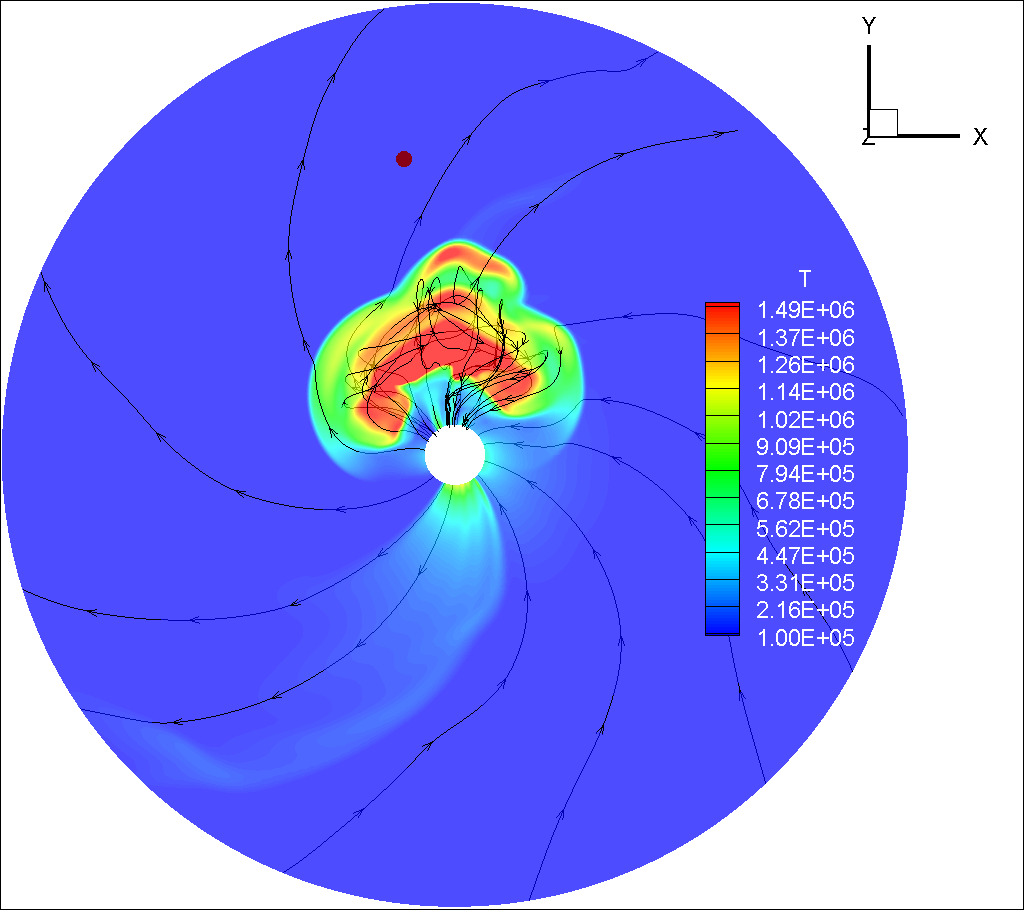}\\
\includegraphics[scale=0.1,angle=0,width=6cm,keepaspectratio]{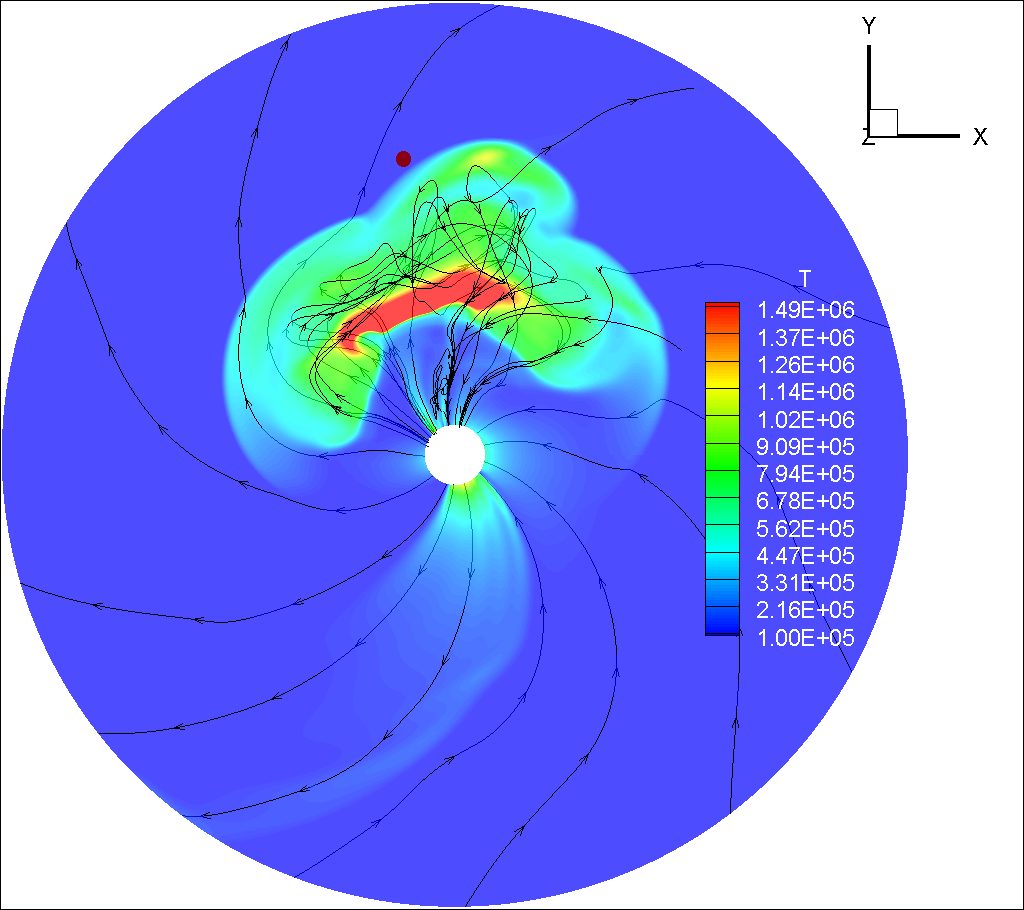}
\includegraphics[scale=0.1,angle=0,width=6cm,keepaspectratio]{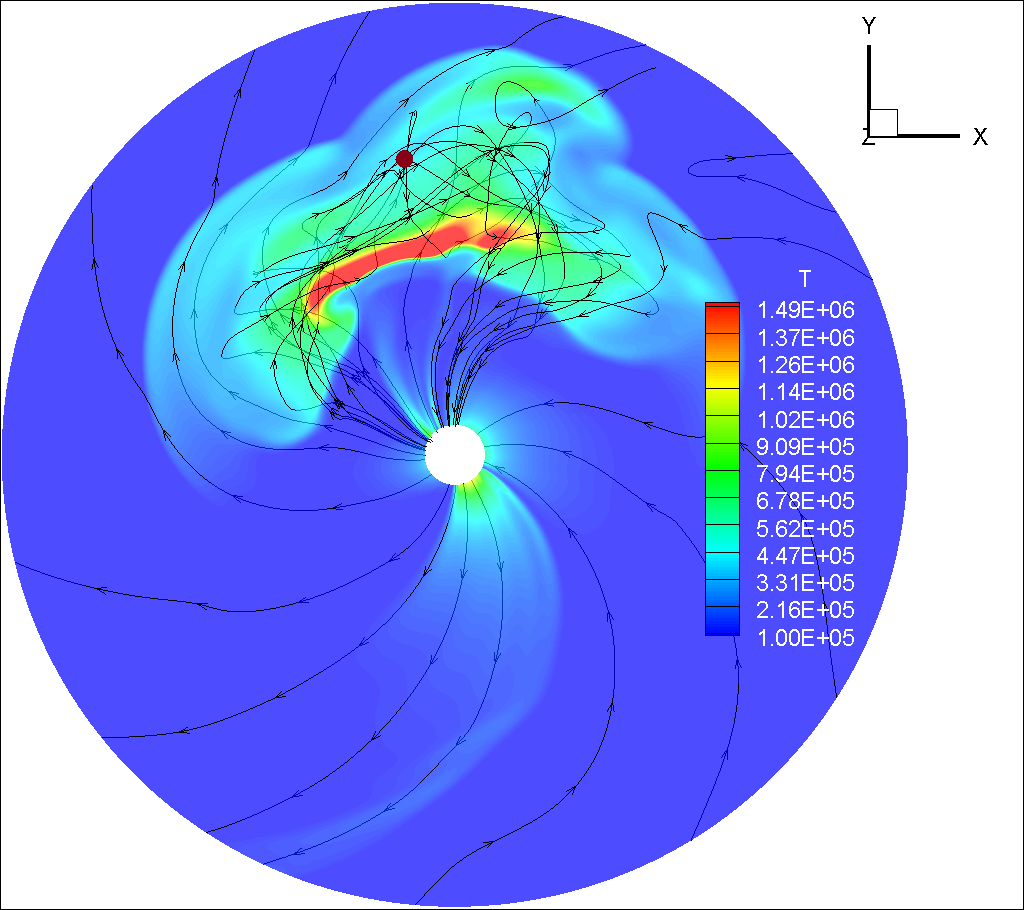}\\
\end{tabular}
\caption{Evolution of the flux rope is shown together with  magnetic field lines in the equatorial slices colored by temperature. The plots are shown 4, 22, 41, and 61 hours after the initial insertion.  {The small brown circle represents a probe at 1 AU in the CME's initial direction. The values probed here are presented in Fig. \ref{Case1_1AU}.}}
\label{CME_evolve}
\end{figure}

The modified spheromak flux rope, when inserted into the solar wind background as described in Sec. \ref{FRinSW}, erupts immediately due to the outwards forces. The eruption of a flux rope shown in Fig. \ref{FR_model} is demonstrated in Fig. \ref{CME_evolve}. This flux rope was inserted into the domain on 12-July-2012 21:30 UT physical time. We can see both the flux rope expansion and formation of a shock in front of the CME, marked by temperature enhancement. We notice that the shock evolution deviates from the initial nearly spherical shape due to the nonuniform solar wind background. The shock front, as well as the flux rope, protrude along the region of high-speed solar wind seen clearly in this direction (Fig. \ref{IH_back}). The shock also changes the direction of the solar wind magnetic field, downstream to it. The speed of the flux rope in the propagation direction was found to be 1172 Km/s. The shock speed, when it reached 50 $R_\odot$ height, was 1309 Km/s. We find these values by fitting a quadratic function to the height-time profile up to 70 $R_\odot$.

It should be noted that the initially inserted flux rope detaches from the inner boundary and its legs experience reconnection with the background solar wind magnetic field.  {Since we introduce the flux rope inside the computational domain, the boundary conditions, which reside in the ghost cells, are not affected by it. Moreover, the solar wind at the inner boundary is already moving faster than the local fast magnetosonic speed. Therefore, no MHD waves can travel back towards the inner boundary.} 

\begin{figure}[!htb]
\center
\includegraphics[scale=0.1,angle=0,width=9cm,keepaspectratio]{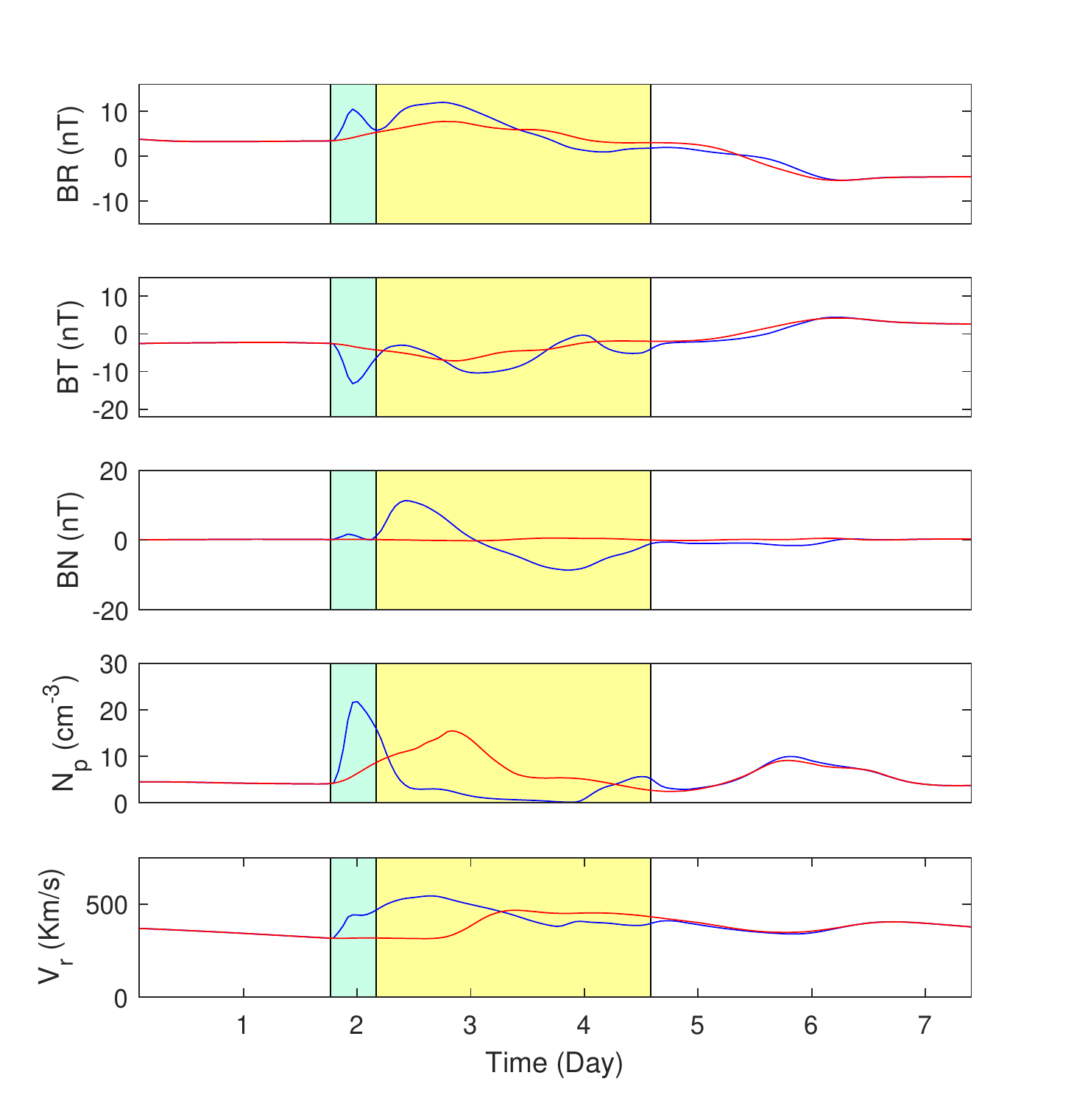}
\caption{1 AU probe results for the flux rope initiated with parameters described in Sec.~\ref{FRinSW} (blue line). The probe is kept in the equatorial plane in the direction of CME initiation. The red line shows the probed data before the CME was launched. The light blue region represents the sheath region, marked by jumps in speed and density. $B_R$ and $B_T$ are also enhanced in this region due to the shock compression. The yellow region marks the flux rope behind the sheath region, identified by non-zero and rotating $B_N$ values. This region is also characterized by very low density, in agreement with observations. The horizontal axis represents time since CME insertion.}
\label{Case1_1AU}
\end{figure}

In Fig. \ref{Case1_1AU}, we show the 1 AU simulation data for this CME. The probe, shown by brown circle, is kept at 1 AU in the direction of CME insertion in the equitorial plane, and the solution is probed every hour at this point. In Fig. \ref{Case1_1AU}, we can clearly see the passage of shock and the flux rope at the probe. We have marked the shock along with its sheath and the flux rope region with blue and yellow color, respectively. The shock and its sheath reveals themselves as abrupt enhancement in $N_p$ and $V_r$ and compression of magnetic field. To identify the end of the sheath region and start of the flux rope, we use the probed $B_N$ values. In the equatorial plane, our inner heliosphere background model gives $B_N \approx 0$. Therefore, the smooth change in the $B_N$ must be due to the flux rope. This region is also marked by a decrease in plasma density, also commonly observed in in-situ data. 

\subsection{Parametric study}
To understand the dependence of CME evolution on its input parameters better, we performed a parametric study, in which we vary some governing parameters keeping the rest of them constant and determining their impact on speed and acceleration of the erupting CME. We simulate 12 cases. The simulation results are given in 
Table~\ref{table}. We also probe the CME properties at 1 AU for these cases. We focus on the following parameters in this study:
\begin{enumerate}
    \item Poloidal flux
    \item Toroidal flux
    \item Initial size of flux rope
    \item The energy multiplier $\xi$
\end{enumerate}
The results are also discussed below in the corresponding subsections.

\begin{table}[!htb]
\center
\begin{tabular}{|c|c|c|c|c|c|c|c|c|}
\hline
Case & $r_0 (R_\odot)$ & \pbox{1.5cm}{Poloidal Flux $\times 10^{21}$Mx}  & \pbox{1.5cm}{Toroidal Flux  $\times 10^{21}$Mx} & \pbox{1.6cm}{Energy \\ Multiplier} & \pbox{1.3cm}{Shock Speed ($\textup{km/s}$)} & \pbox{1.5cm}{Shock accel. ($\textup{km/s}^2$)} & \pbox{1.3cm}{FR speed ($\textup{km/s}$)} & \pbox{1.5cm}{FR accel. ($\textup{km/s}^2$)}  \\ \hline
1    & 15 & 20      & 5   & 2    & 1309  & -0.008   & 1172     & -0.003 \\ \hline
2    & 15 & 15      & 5   & 2    & 1022  & -0.008   & 891      &  0.007 \\ \hline
3    & 15 & 10      & 5   & 2    & 675   &  0.011   & 659      &  0.015 \\ \hline
4    & 15 & 25      & 5   & 2    & 1667  & -0.029   & 1464     & -0.017 \\ \hline
5    & 15 & 20      & 8   & 2    & 1378  & -0.015   & 1175     & -0.010 \\ \hline
6    & 15 & 20      & 10  & 2    & 1362  & -0.016   & 1173     & -0.009 \\ \hline
7    & 15 & 20      & 12  & 2    & 1401  & -0.021   & 1216     & -0.004 \\ \hline
8    & 10 & 20      & 5   & 2    & 1535  & -0.019   & 1379     & -0.018 \\ \hline
9    & 20 & 20      & 5   & 2    & 1044  &  0.005   & 943      &  0.008 \\ \hline
10   & 15 & 20      & 5   & 4    & 1819  & -0.025   & 1629     & -0.028 \\ \hline
11   & 15 & 20      & 5   & 6    & 2162  & -0.036   & 1962     & -0.043 \\ \hline
12   & 15 & 20      & 5   & 1  & 1033  & -0.003   & 937      &  0.005 \\ \hline
\end{tabular}
\caption{Shock and flux rope speeds, and acceleration at 50 $R_\odot$ are shown for the runs with varying initial parameters of the flux rope. The height-time evolution of the CME up to 70 $R_\odot$ is fitted with a quadratic fit to obtain these quantities at 50 $R_\odot$.}
\label{table}
\end{table}

\subsubsection{Relationship between CME speed and acceleration}
\begin{figure}[!htb]
\center
\includegraphics[scale=0.1,angle=0,width=7.5cm,keepaspectratio]{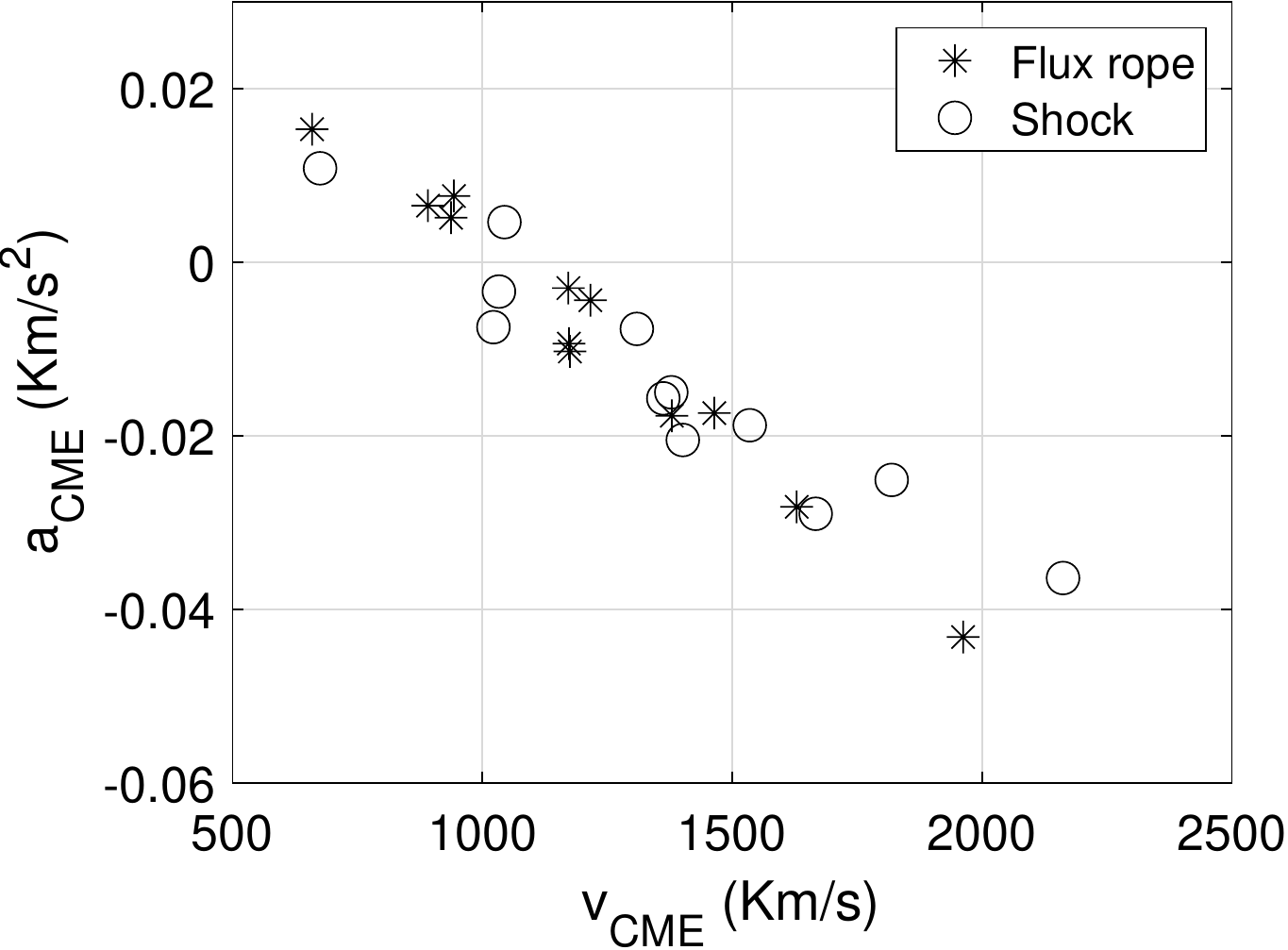}
\caption{Dependence of CME acceleration on its speed at 50 $R_\odot$ for different runs tabulated in Table \ref{table}. Fast CMEs have negative acceleration, whereas the slower ones have positive acceleration, with a clear negative correlation between speed and acceleration.}
\label{a_vs_v}
\end{figure}

Observations show that the background solar wind accelerates slow CMEs and decelerates fast ones. For example,  \citet{Gopal00} show that the average acceleration of CMEs between Solar and Heliospheric Observatory (SOHO) coronagraph field of view and 1 AU is positive for CMEs with initial speed less than 405 km/s and negative for CMEs with initial speed greater than 405 km/s, which is almost equal to the average solar wind speed. To identify this effect in our study, we find the flux rope and shock speeds, and acceleration at 50 $R_\odot$ for all cases. This was done by fitting a quadratic function to height-time data. Figure \ref{a_vs_v} shows the inverse dependence of the flux-rope and shock acceleration on the CME speed, in agreement with observational data. The speed at which the acceleration changes from positive to negative is around 1000 km/s, which is higher than the ambient solar wind speed. This is stipulated by the method we use for the CME insertion. It results in large acceleration initially, when a CME is just inserted into the solar wind. This is why, acceleration dominates over the solar wind drag at 50 $R_\odot$, where these data are taken.  

\subsubsection{Effect of Poloidal flux on CME evolution}
\begin{figure}[!htb]
\center
\includegraphics[scale=0.1,angle=0,width=7.5cm,keepaspectratio]{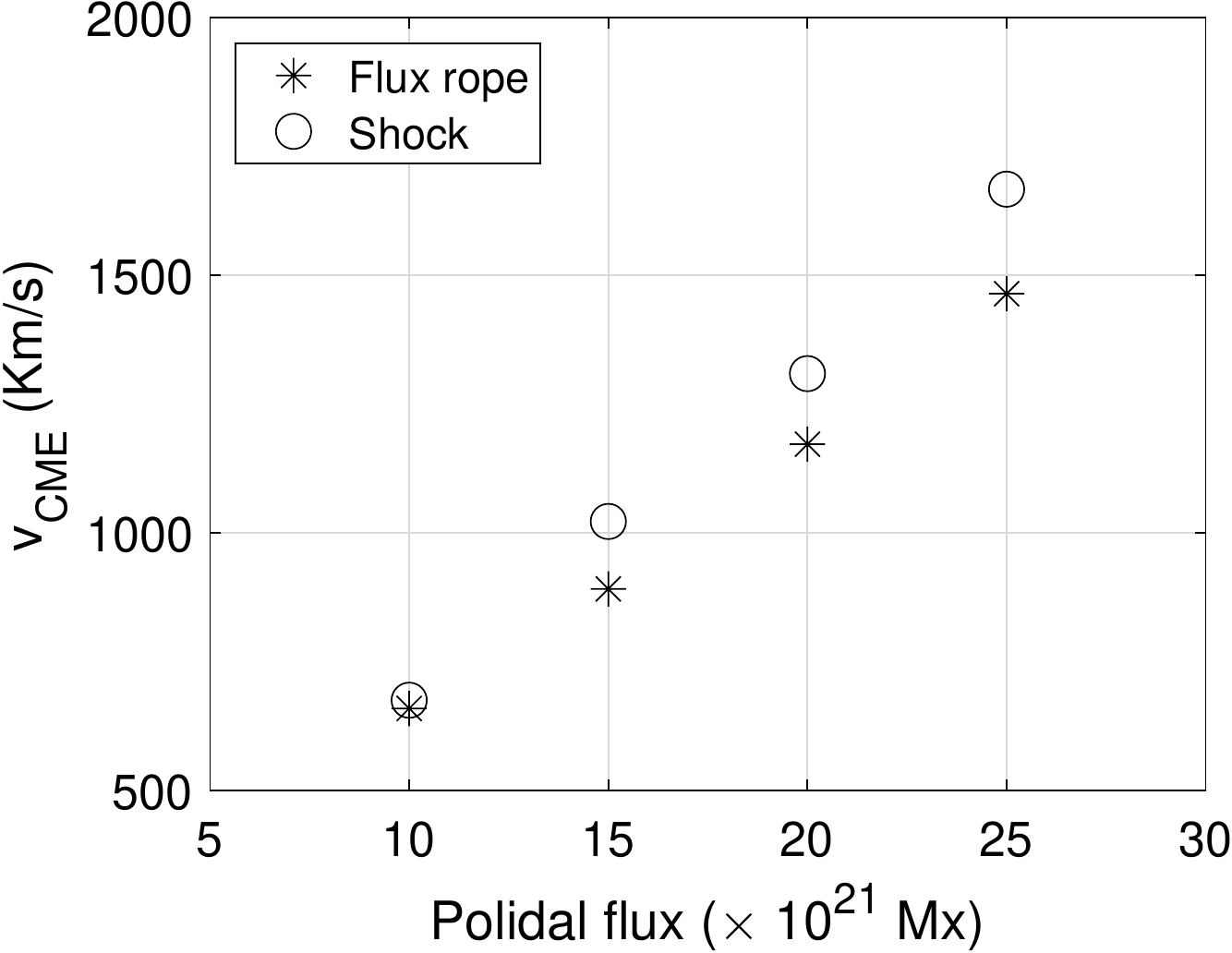}
\caption{Dependence of the simulated CME speed at 50 $R_\odot$ on the corresponding input poloidal flux. We find a linear trend relating these two properties.}
\label{pol_vs_v}
\end{figure}

Keeping $r_0=15 R_\odot$, $r_1=26 R_\odot$, $a=5 R_\odot$, $\xi = 2$, and the toroidal flux = $5\times 10^{21}$ Mx, we vary the poloidal flux within 10--25 $\times 10^{21}$ Mx. The resulting flux-rope and shock speeds are shown in Fig. \ref{pol_vs_v}. We see a strong linear dependence of the CME speed on the poloidal flux. A similar result was reported by \citet{Singh19a} and \citet{Jin17} using the Gibson-Low flux rope model.  {This relationship is observed because by increasing the poloidal flux in a flux rope, we are increasing the magnetic pressure and tension forces responsible for the eruption. We will discuss this point in more detail in the next subsection.} A positive correlation between CME speeds and corresponding poloidal fluxes was demonstrated earlier in observations by \citet{Gopalswamy18}.

\begin{figure}[!htb]
\center
\includegraphics[scale=0.1,angle=0,width=9cm,keepaspectratio]{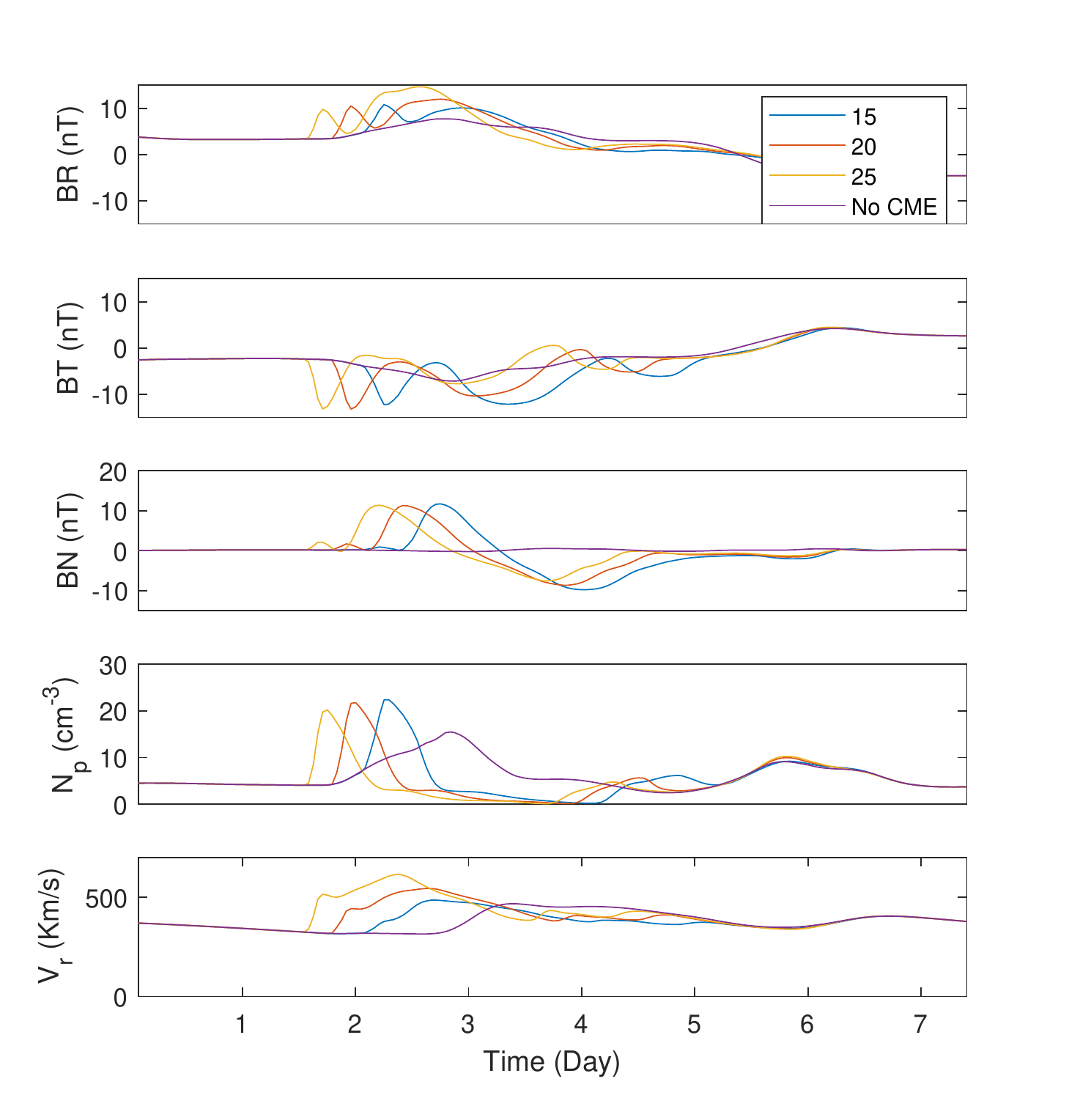}
\caption{1 AU probe data for cases with different input poloidal flux. The probe position in the equatorial plane and aligned with the CME launch direction. The legend shows the input poloidal flux in  Mx/$\times 10^{21}$. The horizontal axis represents time since CME insertion.}
\label{pol_1AU}
\end{figure}

In Fig. \ref{pol_1AU}, we show the properties of the simulated CME at 1 AU in the equatorial plane and in the direction of CME launch. Here we show the CMEs with poloidal flux of $15\times10^{21}$, $20\times10^{21}$, and $25\times10^{21}$ Mx, along with the case when no CME was launched. The magnetic field is given in RTN coordinates, where the $B_N$ values are approximately equal to the $B_z$ values in the geocentric solar ecliptic (GSE) coordinates. The shock arrival is characterized  by a sudden increase in density and radial speed. The flux-rope arrival reveals itself by a large change in  $B_N$. The changes in $B_R$ and $B_T$, which occur before the change in $B_N$, are due to the shock compression. The results at 1 AU show that CMEs with higher poloidal flux arrive early because  they are launched at higher speeds. We also see that the passage of a flux rope is accompanied by a drop in density, as often observed in \textit{in situ} data \citep{Burlaga81}. 

\subsubsection{Effect of toroidal flux on the CME evolution}
\begin{figure}[!htb]
\center
\begin{tabular}{c c c}  
\includegraphics[scale=0.1,angle=0,width=5cm,keepaspectratio]{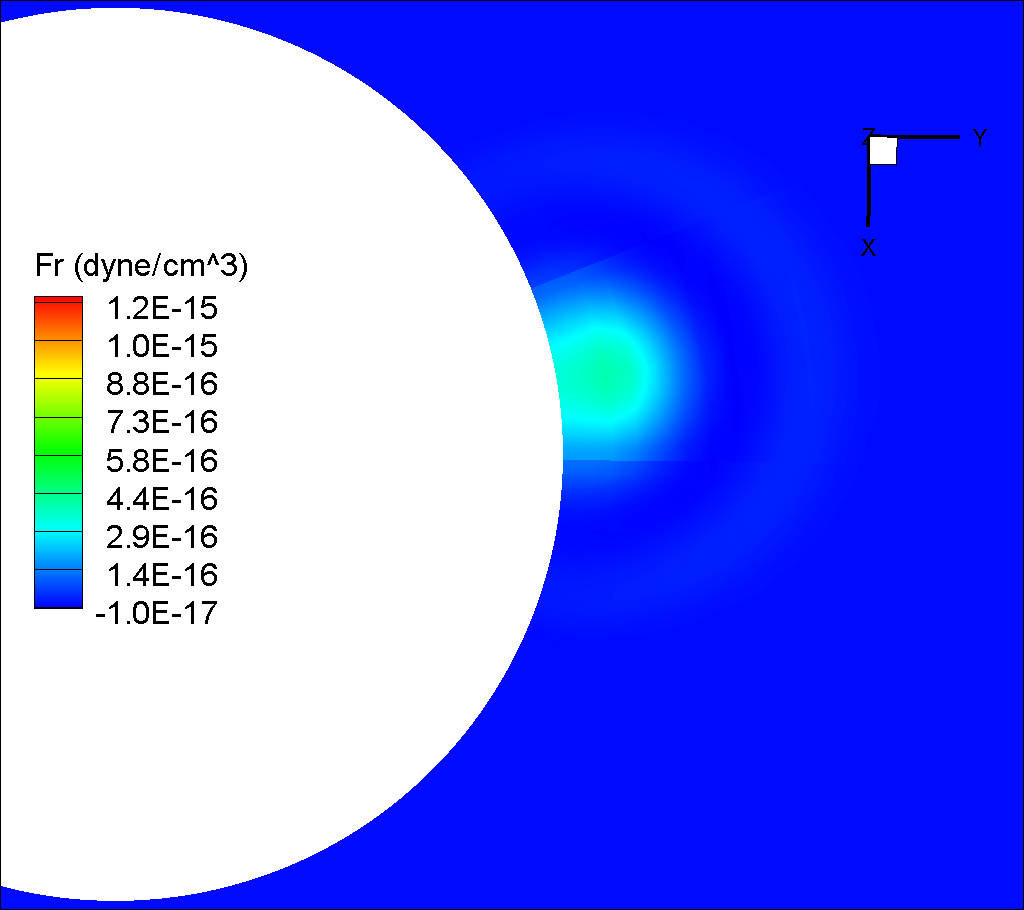}
\includegraphics[scale=0.1,angle=0,width=5cm,keepaspectratio]{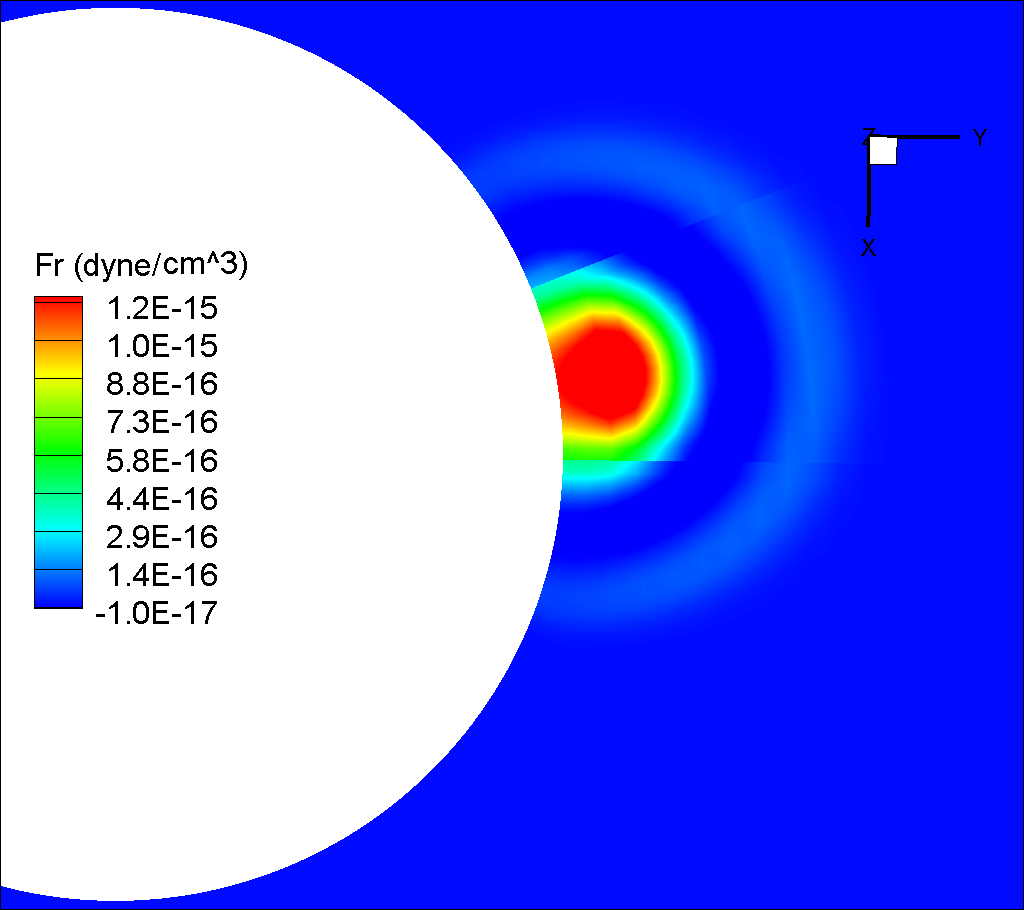}
\includegraphics[scale=0.1,angle=0,width=5cm,keepaspectratio]{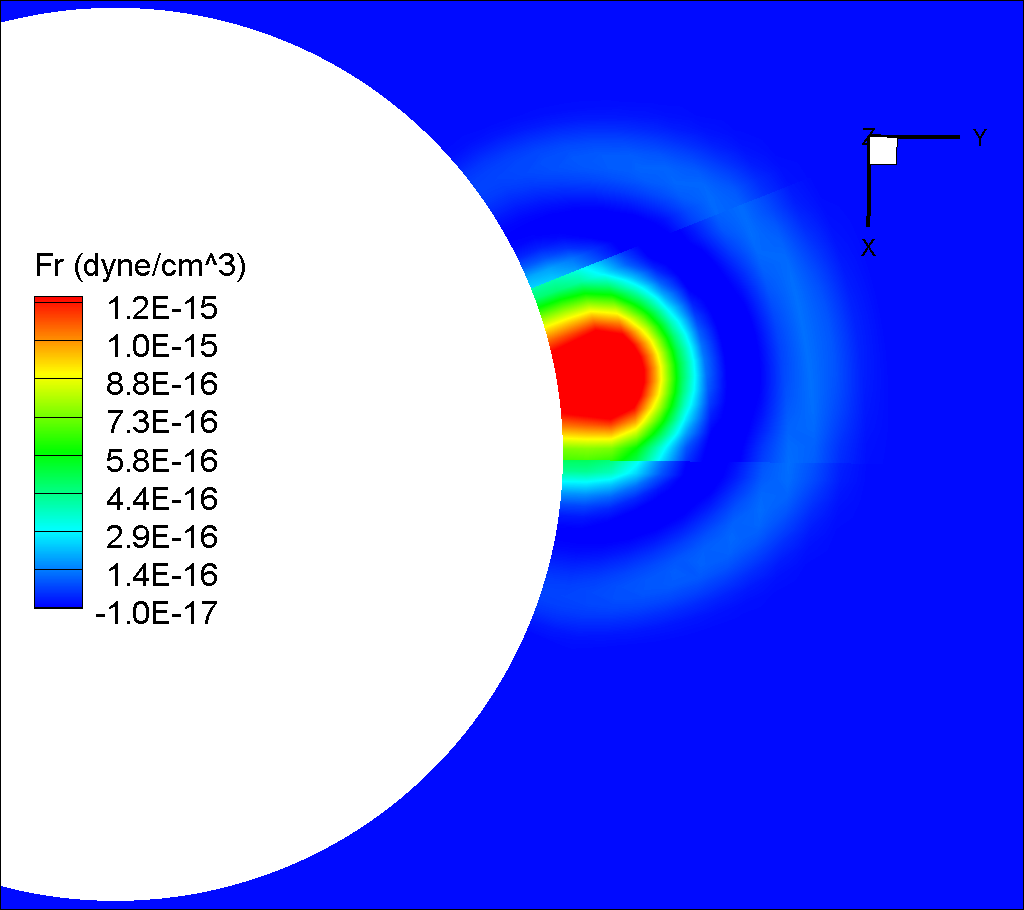}
\end{tabular}
\caption{ {The slices showing radial force distribution in the FRs in the equitorial plane with varying fluxes. The (poloidal,toroidal)$\times 10^{21}$Mx fluxes are (from left to right)  
 (10,5), (20,5), and (20,10)}. We can see that the poloidal flux controls the force distribution much more than the toroidal flux.}
\label{force}
\end{figure}

\begin{figure}[!htb]
\center
\includegraphics[scale=0.1,angle=0,width=7.5cm,keepaspectratio]{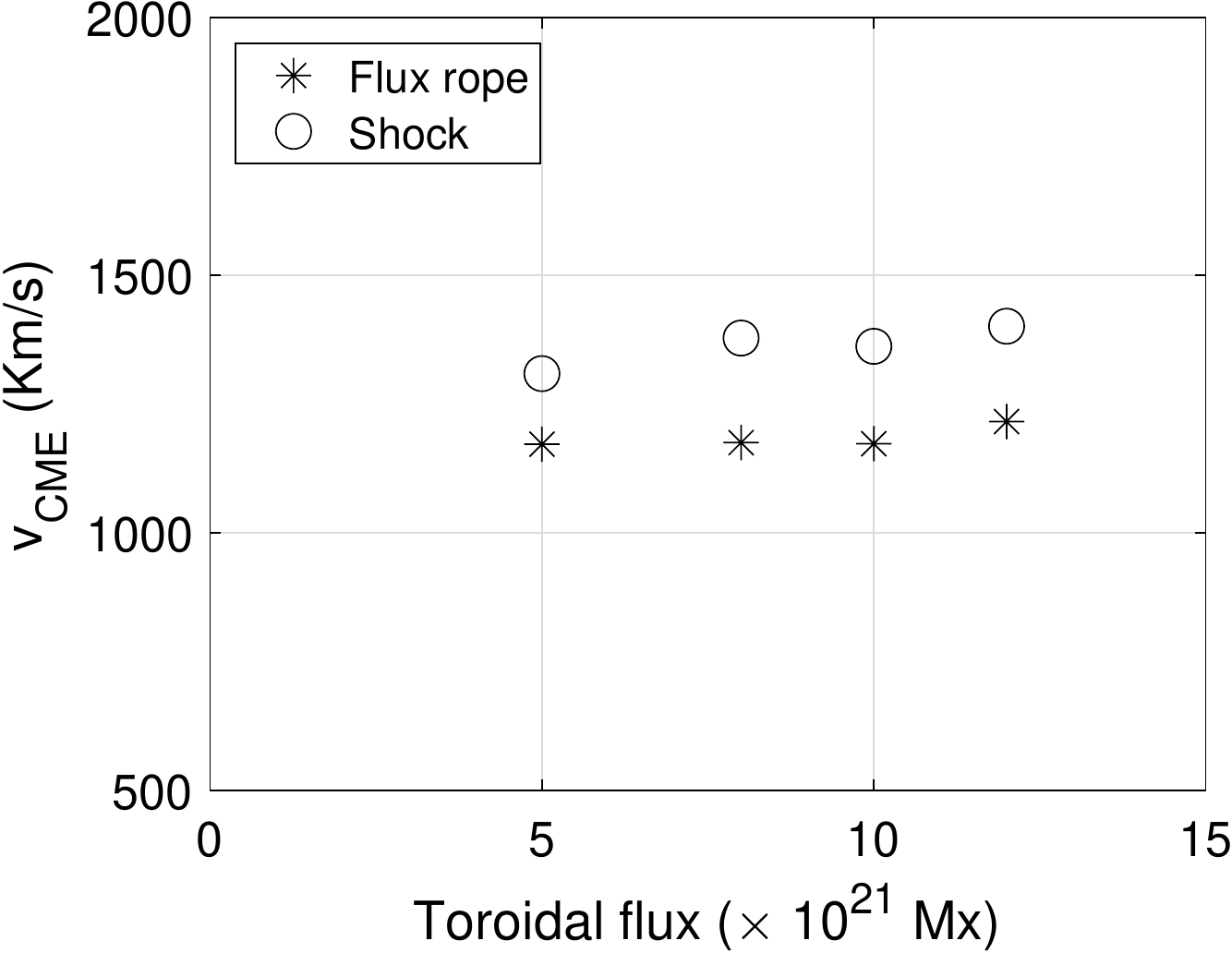}
\caption{Dependence of the simulated CME speed at 50 $R_\odot$ on the input toroidal flux. No linear trend is seen contrary to its dependence on the in poloidal flux.}
\label{tor_vs_v}
\end{figure}

We vary the toroidal flux in the flux rope in the range of 5--12 $\times 10^{21}$ Mx, while keeping $r_0=15 R_\odot$, $r_1=26 R_\odot$, $a=5 R_\odot$, $\xi = 2$, and poloidal flux = $20\times 10^{21}$ Mx. The effect of this on CME speed is plotted in Fig. \ref{tor_vs_v}. Here we see a stark difference in the trend as compared with that of the  variation of the poloidal flux. The CME speed is much less dependent on the input toroidal flux.  {This phenomenon can be understood by looking at the dependence of force distribution inside the flux rope on the varying poloidal and toroidal fluxes. In Fig. \ref{force}, we show the radial force acting on the flux rope at the time of its insertion as calculated by Eqn. \ref{force_eqn}. The three cases shown have the poloidal and toroidal fluxes equal to 10 and 5, 20 and 5, and 20 and 10 respectively. The units used are $10^{21}$Mx. We show that on doubling the poloidal flux from $1\times10^{22}$ Mx to $2\times10^{22}$ Mx while keeping toroidal flux equal to $5\times10^{21}$ Mx, the force increases considerably. However, when we double the toroidal flux from $5\times10^{21}$ Mx to $10\times10^{21}$ Mx while keeping the poloidal flux as $2\times10^{22}$ Mx, there is very little change in the force distribution. Thus we conclude that in a curved flux rope geometry such used in our model, the poloidal flux is the main contributor of the eruptive force, and hence affects the CME speed stronger than the toroidal flux. Since the observed CMEs also have such curved geometries, this statement applies to them as well \citep{Kliem2006}}. Observations show that the majority of poloidal flux in a CME is due to the reconnected flux created during the eruption \citep{Qiu07, Longcope07}. Thus, the major factor that is affecting the CME speed is the magnetic flux which a flux rope gains during its eruption, not the flux it possesses in the pre-eruptive flux rope.

\begin{figure}[!htb]
\center
\includegraphics[scale=0.1,angle=0,width=9cm,keepaspectratio]{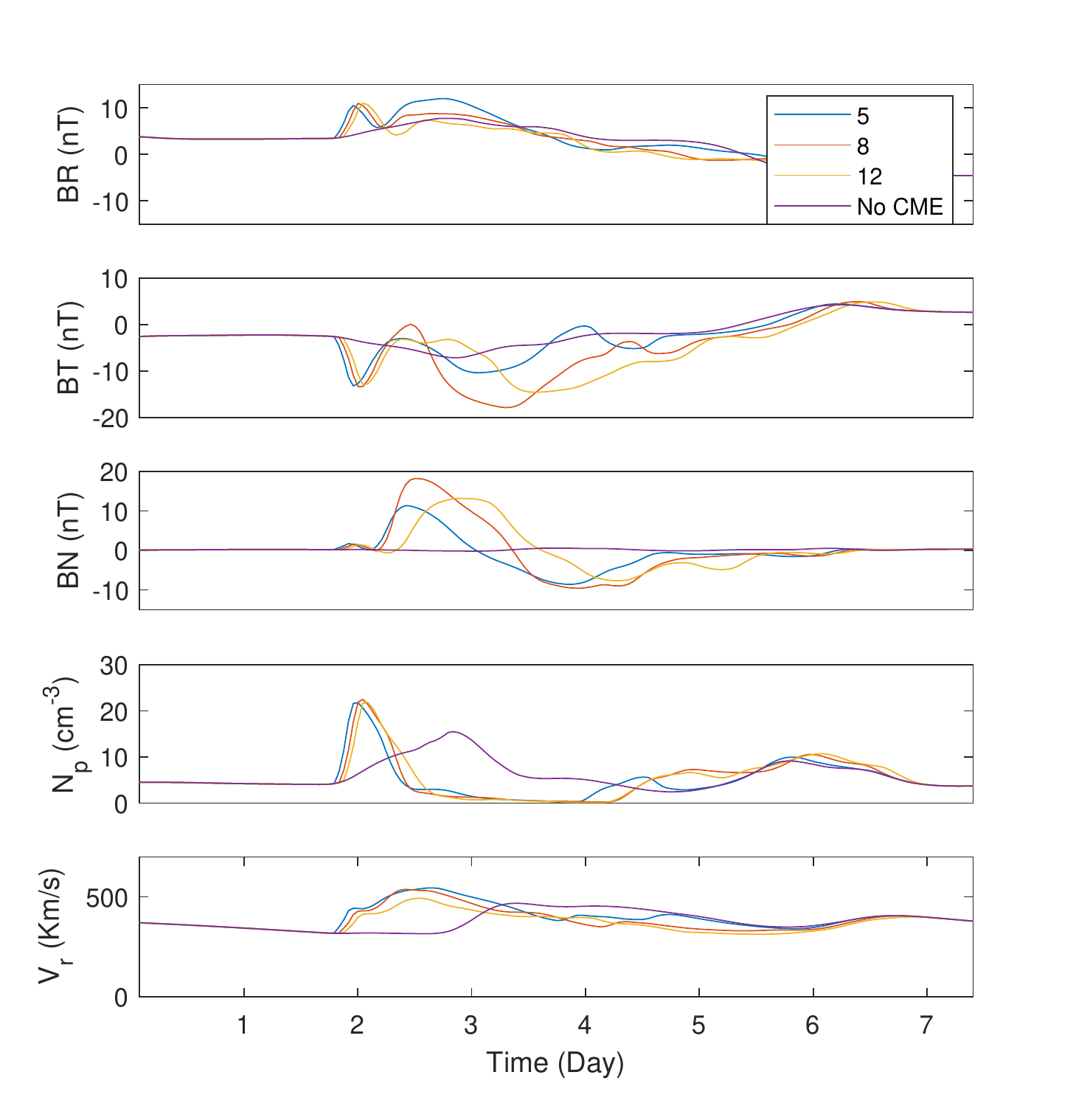}
\caption{Probe data at 1 AU for the cases of different input toroidal flux. The probe is kept in the equatorial plane on the line along the CME launch direction. The legend shows the input toroidal flux in MX/$\times 10^{21}$). The horizontal axis represents time since CME insertion.}
\label{tor_1AU}
\end{figure}

Figure \ref{tor_1AU} shows the probe data at 1 AU for the runs with toroidal flux of $5\times10^{21}$, $8\times10^{21}$, and $12\times10^{21}$ Mx. We see that the arrival times of CME shocks do not differ much, but magnetic field magnitudes inside them are very different. This shows that CMEs with the same poloidal flux, speed, orientation, and direction can carry different magnetic field at 1 AU, and hence have different geo-effectiveness. This means that the  {toroidal} magnetic flux in pre-eruptive flux ropes may not affect the CME speed, but it does impact its features at 1 AU.

\begin{figure}[!htb]
\center
\includegraphics[scale=0.1,angle=0,width=9cm,keepaspectratio]{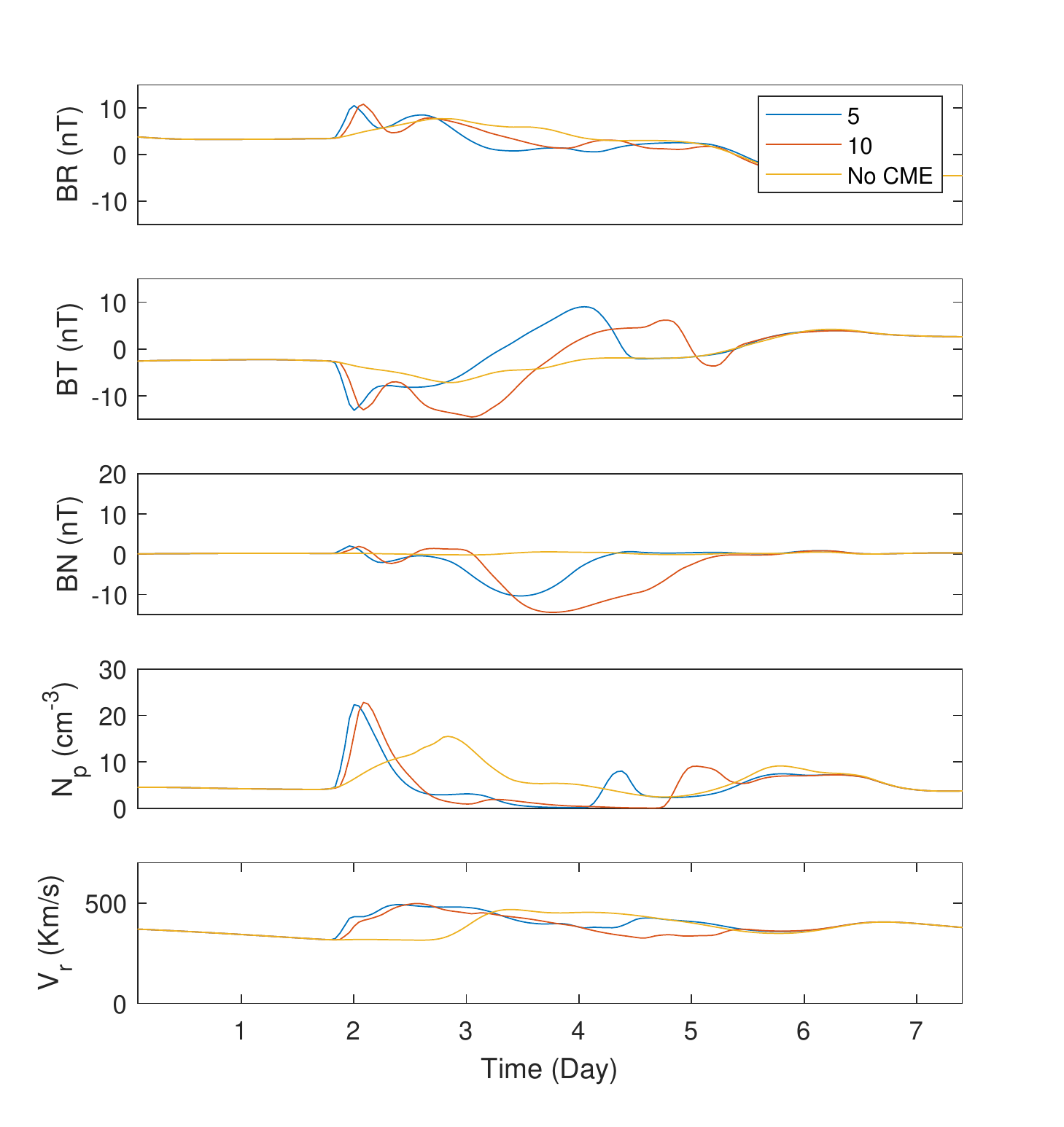}
\caption{Probe data at 1 AU for the cases of different input toroidal flux for the case where the flux rope is tilted by 90 degrees, so that the toroidal flux can contribute to  $B_N$. The legend shows the input toroidal flux in Mx/$\times 10^{21}$. One can see that two CMEs with similar initial speed and  arrival times can have different $B_N$, which affects their geo-effectiveness. The horizontal axis represents time since CME insertion.}
\label{90_tilt}
\end{figure}

To demonstrate this point even further, we made two more simulations with the toroidal flux of $5\times10^{21}$ Mx and $10\times10^{21}$ Mx, the orientation tilted by 90 degrees, and while the rest of parameters are the same. This should create conditions for the CME toroidal flux to contribute to the $B_N$ values at 1 AU, in accordance with the self-similar expansion conditions. Although our simulated CMEs do not expand in a self-similar way, they retain the overall coherent structure, so we should expect the toroidal flux to contribute to $B_N$ in any event. Figure \ref{90_tilt} shows the comparison of these two CMEs at 1 AU. We see that the toroidal flux makes $B_N$ negative. It is worth noticing here is that even though the CMEs arrive at roughly the same time and have the same initial orientation and poloidal flux, the $B_N$ values are much more negative for the case of $10\times10^{21}$ Mx toroidal flux. Thus, this CME is more geoeffective. Therefore, if we wish to use our flux rope-based model for CME predictions, constraining poloidal flux alone is not sufficient. We need to ensure a proper magnitude of toroidal flux as well.


\subsubsection{Effect of total energy on CME evolution}
\begin{figure}[!htb]
\center
\includegraphics[scale=0.1,angle=0,width=7.5cm,keepaspectratio]{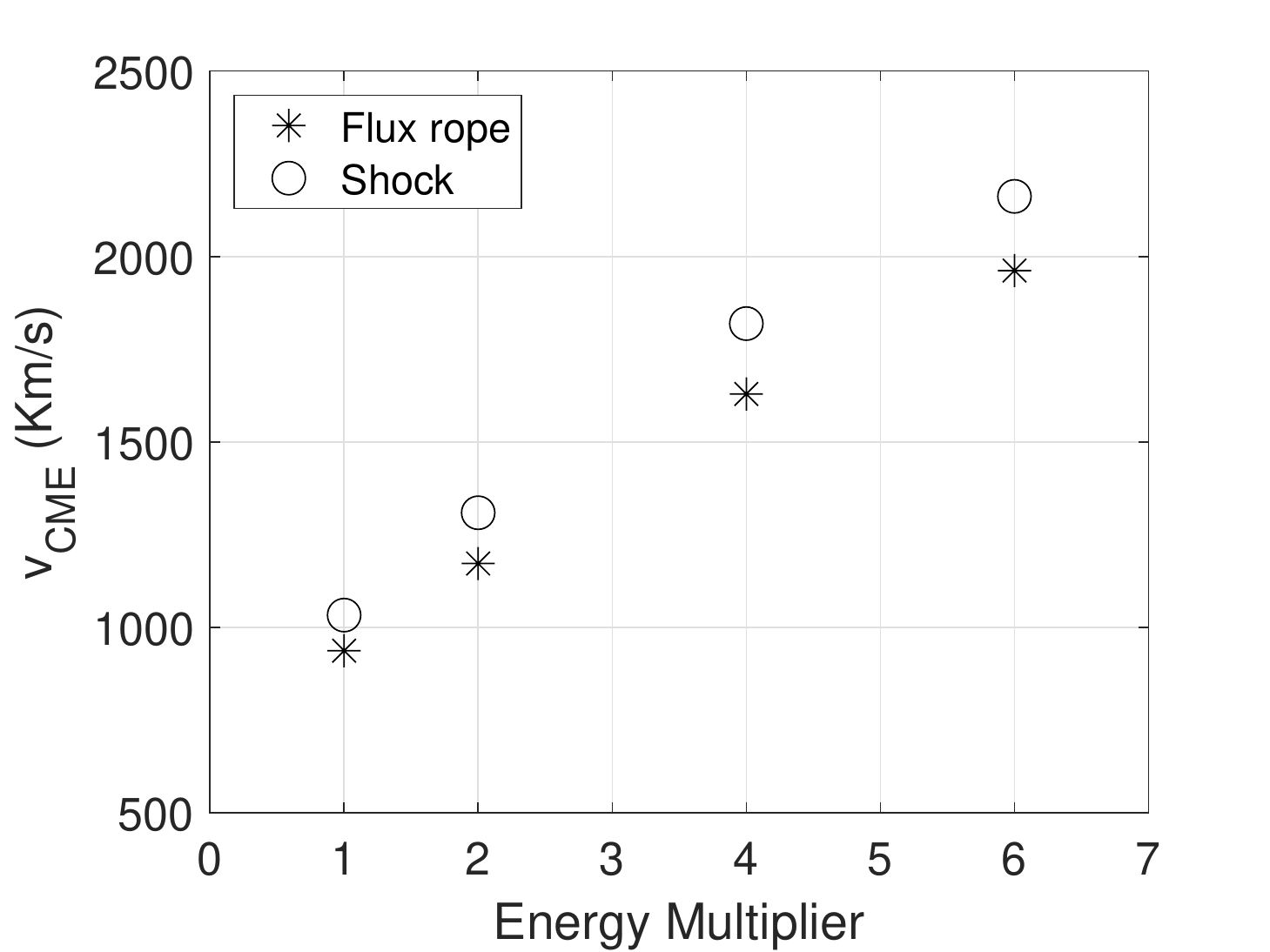}
\caption{CME speed dependence on the total energy multiplier $\xi$. The CME speed is calculated at 50 $R_\odot$ using a quadratic fit to height-time data.}
\label{Emult_vs_v}
\end{figure}

As discussed in Sec. \ref{FRinSW}, we use a multiplier $\xi$ while adding the flux rope total energy to the simulation domain. We find that this multiplier can control the eruption speed of the flux rope and therefore can be used to constrain it to desired values. To quantify the effect of this parameter, we vary it between the values of 1 and 6 while keeping $r_0=15 R_\odot$, $r_1=26 R_\odot$, $a=5 R_\odot$, and the poloidal and toroidal fluxes equal to as $20\times10^{21}$ Mx and $5\times10^{21}$ Mx, respectively. The effect of this parameter on the flux rope and shock speeds at 50 $R_\odot$ is shown in Fig. \ref{Emult_vs_v}. We see a linear dependence of the speed on $\xi$. This relationship can be used to launch CMEs with desirable speed.  {However, we note that according to the relation $\beta=(\gamma-1)(\xi-1)$, the choice of $\xi>3$ results in $\beta>1$ in our model, which is an unnatural property for a CME. Therefore, $\xi=3$ should be considered as the realistic upper limit for our model.}

\begin{figure}[!htb]
\center
\includegraphics[scale=0.1,angle=0,width=9cm,keepaspectratio]{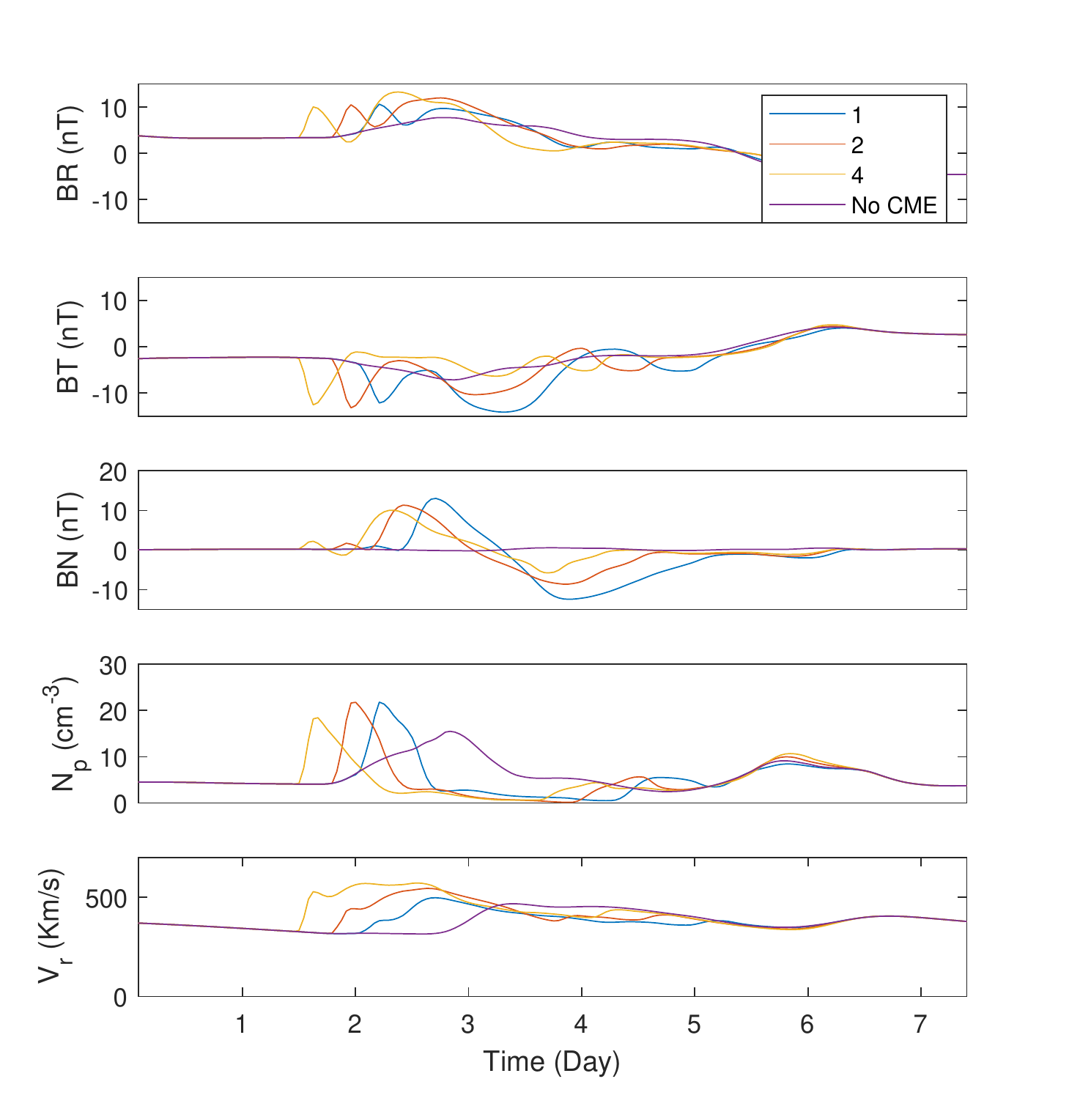}
\caption{1 AU probe data for different input energy multipliers $\xi$. The legend shows the energy multiplier used in simulation. The probe is kept in the equatorial plane in the CME launch direction. The horizontal axis represents time since CME insertion.}
\label{Emult_1AU}
\end{figure}

\subsubsection{Effect of the initial size a flux rope on the CME evolution}
\begin{figure}[!htb]
\center
\includegraphics[scale=0.1,angle=0,width=7.5cm,keepaspectratio]{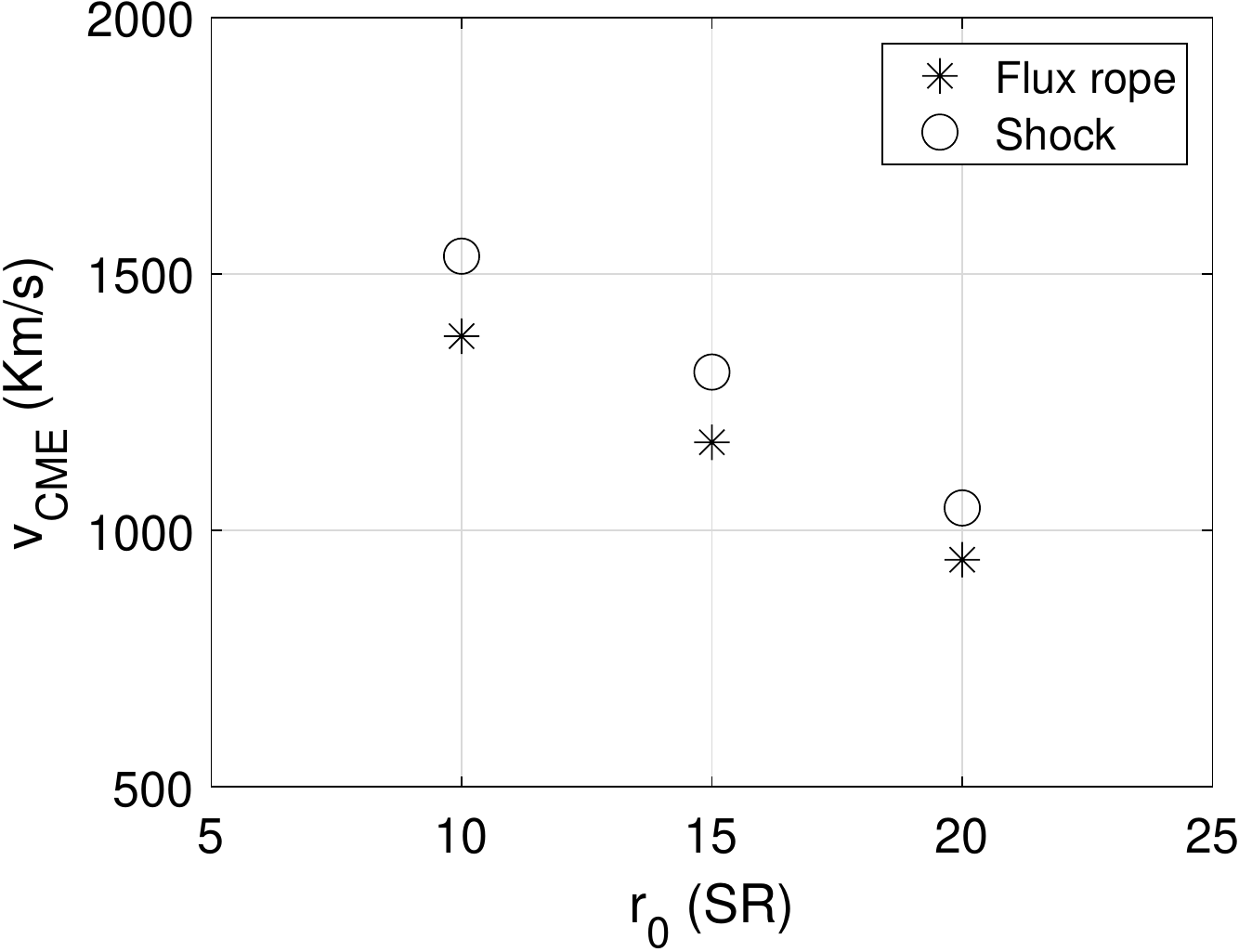}
\caption{We show the dependence of the simulated CME speed at 50 $R_\odot$ on the initial size of the flux rope, $r_0$. A negative trend is seen, which is similar to the one obtained in the solar corona  \citet{Singh19a}.}
\label{r0_vs_v}
\end{figure}

\begin{figure}[!htb]
\center
\includegraphics[scale=0.1,angle=0,width=9cm,keepaspectratio]{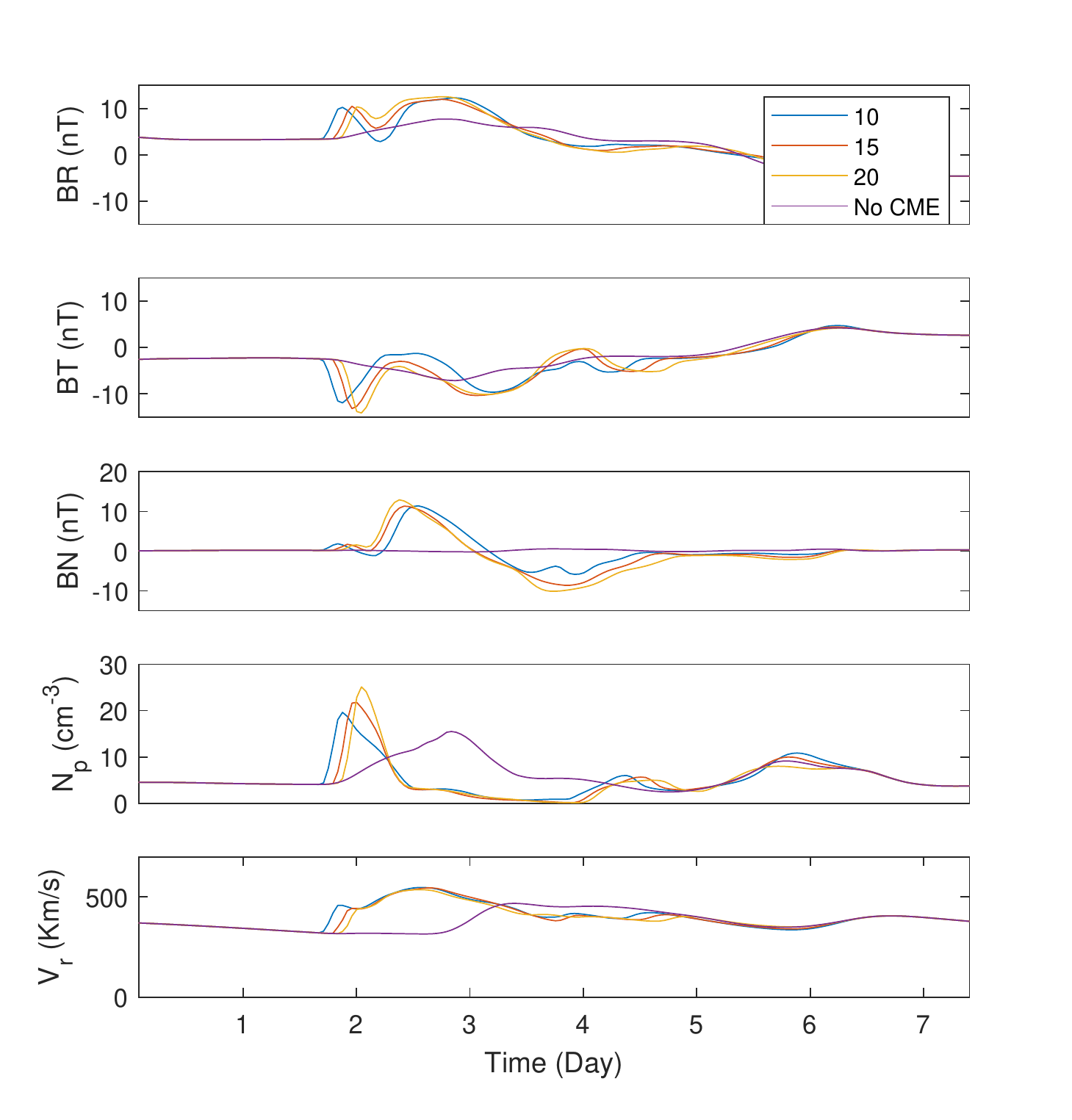}
\caption{Earth probe data for the cases with different input initial size parameter $r_0$. The legend shows the values of $r_0$ used in our simulations, scaled to $R_\odot$. The horizontal axis represents time since CME insertion.}
\label{r0_1AU}
\end{figure}

If we want to change the initial size of a flux rope, while keeping the magnetic flux values unchanged, we need to modify the magnetic field strength in this flux rope. In this way, a bigger initial flux rope will result in a smaller magnetic field strength inside it and decrease the total pressure. This means that larger initial flux rope will erupt at a smaller speed. This effect was also shown in the CME simulations in the solar corona using a modified spheromak model  \citep{Singh20}. We show that this also takes place in our inner-heliosphere simulations by keeping $r_1=26 R_\odot$, $a=5 R_\odot$, poloidal and toroidal fluxes equal to $20\times10^{21}$ Mx and $5\times10^{21}$ Mx, respectively, $\xi=2$, while varying $r_0$ in the range of 10--20 $R_\odot$. The results are shown in Fig. \ref{r0_vs_v}. We see a reduction in both shock and flux rope speeds with increasing $r_0$. The Earth probe data for these runs are given in Fig. \ref{r0_1AU}. 

\subsection{CME-CME collision}\label{sec_coll}
During solar maxima, CME-CME collision scenarios are common due to a large number of CMEs erupting from the Sun. Such collisions are possible if a slower CME is in the path of a faster one. The resulting dynamics of the CMEs can range from purely inelastic to superelastic \citep[][and references therein]{Shen17}. A collision can change the CME direction, as well as its speed, resulting in the CME features observed at 1 AU totally different from such with a single CME  {\citep[][and references therein]{Manchester2017}}. This makes the collision process very relevant for space weather predictions. A CME model should be able to simulate CME-CME interactions for it to be operationally viable. Flux ropes of each CME play an integral part in the collision process. This is why, we expect flux-rope-based models to be suitable for such simulations. Many authors have shown the ability of different flux rope models to simulate CME-CME collision process \citep[eg.][]{Lugaz05, Shen16, Shiota16}.

\begin{figure}[!htb]
\center
\includegraphics[scale=0.1,angle=0,width=12cm,keepaspectratio]{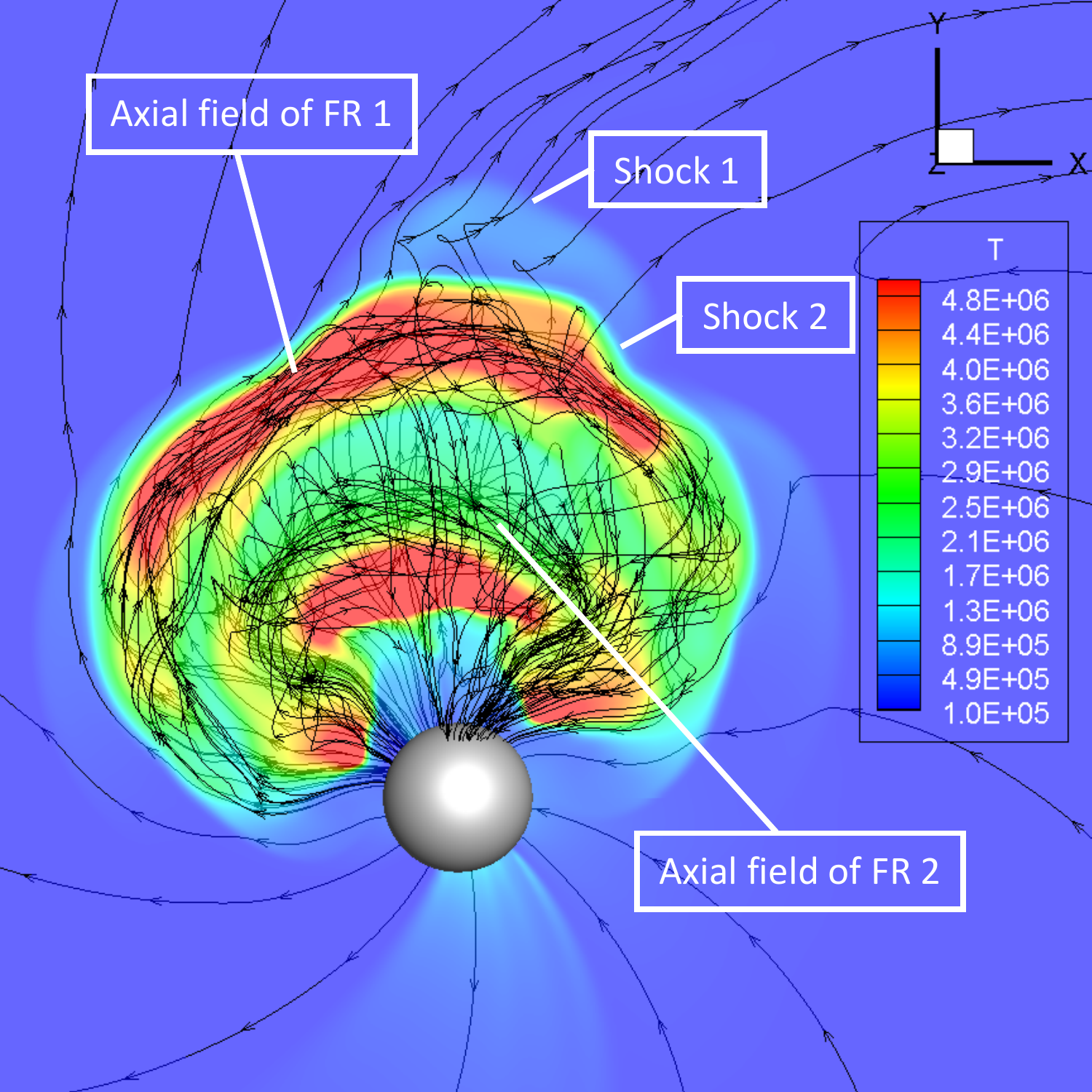}
\caption{Colliding CMEs discussed in Sec. \ref{sec_coll}, 33 hrs after the launch of the slower CME. We show the equatorial plane colored by plasma temperature along with the magnetic field lines. The axial field lines belonging to both CMEs can be seen. The poloidal field lines  become very complicated, owing to the magnetic reconnection between them. The shock of the faster CME, labeled here as ``Shock 2" has crossed through the slower CME's flux rope by this time.  {The image size is 310$\times$310 $R_\odot$}.}
\label{Collision}
\end{figure}

To show the applicability of the modified spheromak model to simulate CME-CME interaction, we launch 2 CMEs, with case numbers 11 and 12 in Table \ref{table}. The slower CME was launched as in case 12 and had shock speed of 1033 km/s at 50 $R_\odot$. When this CME shock reaches 125 $R_\odot$, 19 hrs after the insertion, a faster CME with parameters of case 11 is launched in the same direction. The flux rope of the slower CME is at 116 $R_\odot$ at this time. Owing to its larger speed, this CME catches up the slower CME and the collision process begins. In Fig. \ref{Collision}, we show these two CMEs during the collision phase, 33 hrs after the launch of the slower CME. We can see the axial field lines of two flux ropes and a complex structure made by the poloidal field lines. The shock in front of the slower CME is at 180 $R_\odot$, its flux rope being at 172 $R_\odot$ at this time. The shock due to the faster CME has already crossed most of the flux rope of the slower CME by this time. 

\begin{figure}[!htb]
\center
\includegraphics[scale=0.1,angle=0,width=9cm,keepaspectratio]{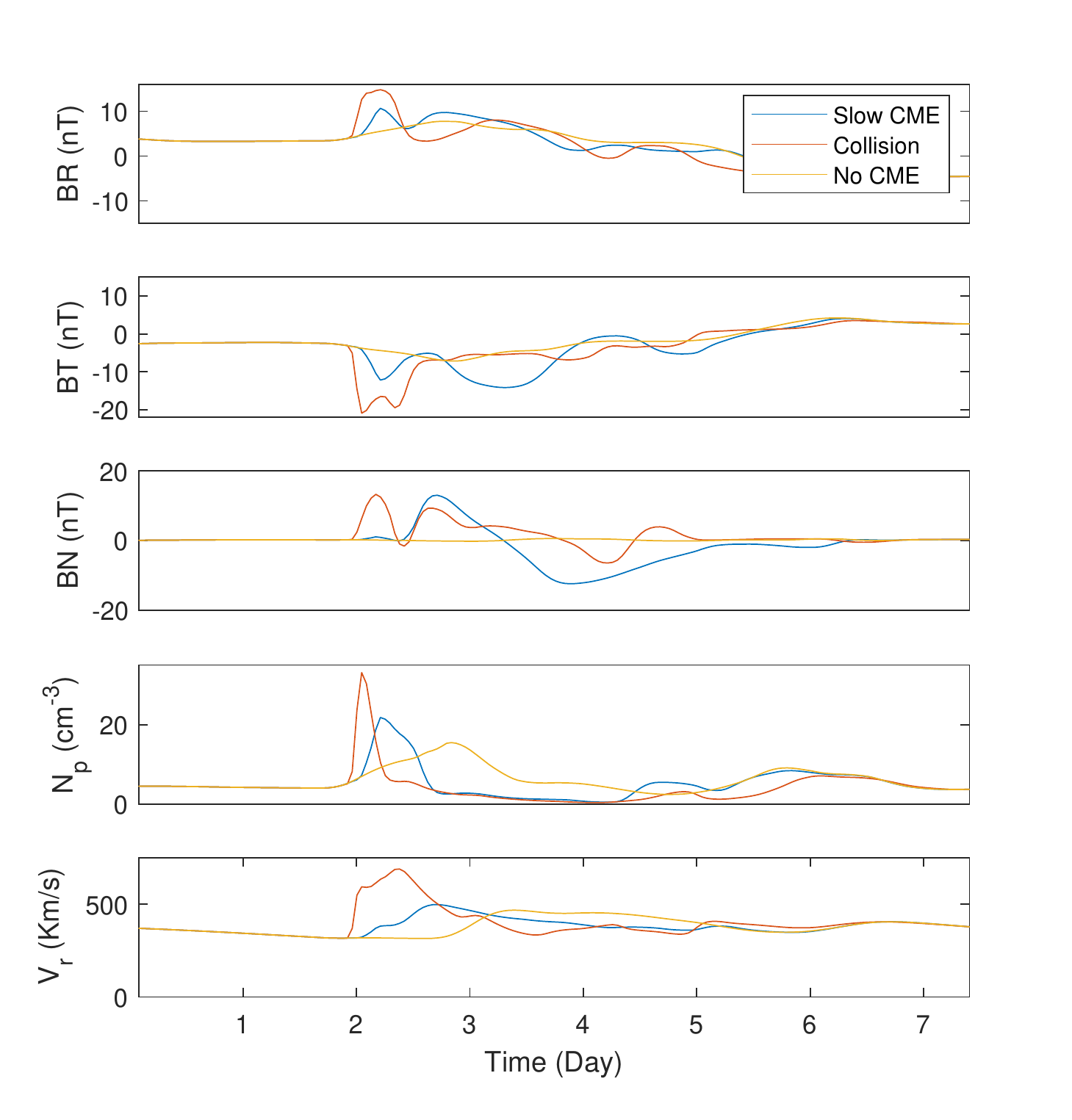}
\caption{Earth probe data from the collision simulations are shown together with the solution for a slow CME only. We find that the collision resulted in the earlier arrival of slow CME. The horizontal axis represents time since CME insertion.}
\label{Collision_1AU}
\end{figure}

We show the Earth probe data for this collision in Fig. \ref{Collision_1AU}, together with the solution involving only first, slower CME. We find that the shock reaches the probe at the same time in both cases, but the flux rope of the slower CME reaches 1 AU faster when no CME is pushing it from behind. We also find that the density jump, as well as the bulk speed in the case of collision, is stronger. In our example, the orientation of two colliding flux ropes was such that the field lines behind the slow CME and ahead of the faster CME had opposite signs in the $z$ direction. This resulted in magnetic reconnection of these field lines, greatly reducing the magnetic field strength in the $z$ direction. This can be seen as a reduction in $B_N$ in the case of collision (see Fig.~\ref{Collision_1AU}, once again confirming the fact that CME collisions can change the geo-effectiveness of CMEs.

\subsection{12 July 2012 CME}\label{sec_12July}
In this section, we investigate the applicability of our model to simulations of actual CMEs. We have chosen the 12 July 2012 CME for this purpose. The choice was made for several reasons. The speed, direction, and orientation of this CME could be reliably obtained from three different viewpoints of the Solar Terrestrial Relations Observatory (STEREO) and Solar and Heliospheric Observatory (SOHO) spacecraft, since STEREO A and B were making an almost right angle to the Sun-Earth line during this time. This also means that the projection effects can be removed from the mass estimates with higher accuracy. The poloidal and toroidal fluxes were calculated also more accurately, since the source of this CME was near  {the solar disk center as viewed from the Earth} and the solar magnetograms, used to determine these values, are most accurate in this region. \citet{Singh20} have explained the method of determination of these values in detail, using the 12 July 2012 case. They find that this CME had the following properties:
\begin{enumerate}
\item The CME speed was found to be 1265 Km/s using a linear fit to the height-time profile.
\item The direction of the CME was found to be at 12 degrees south and 8 degrees west. 
\item The CME flux rope orientation, found using the GCS method, was 53 degrees with respect to the solar equator.
\item The poloidal flux of the CME was found to be $1.4\times10^{22}$ Mx.
\item The toroidal flux of the CME was found to be $2.1\times 10^{21}$ Mx.
\item The CME flux rope was found to have a positive helicity sign.
\item The mass of the CME was found to be $1.65\times 10^{16}$ g.
\end{enumerate}

As explained in Sec. \ref{models}, we can input all these values easily into our model. 

There is one outstanding problem though. Our model shows a positive acceleration at the initial insertion, which can be at a height of 30-40 $R_\odot$. But the actual CME can be experiencing deceleration at those heights. We somehow need to match the speeds of the simulated and observed CMEs at some height after the initial insertion. In this study, we match the speed of the simulated CME at 50 $R_\odot$ with the speed of observed CME at the same height obtained using the drag based model (DBM) \citep{Vrsnak07}. Using $V_{CME}= 1265$ at 15 $R_\odot$, the drag parameter of $0.1\times10^{-7}$ and the asymptotic solar wind speed equal to 400 km/s, BDM gives the CME speed at 50 $R_\odot$ equal to 1140 km/s. The DBM model also predicts the CME height on 12-July-2012 21:30 UT to be 34 $R_\odot$. The drag parameter and the asymptotic solar wind speed values used here are based on the recommendation by \citet{Vrsnak14}. Using $r_0 = 15$, $r_1 = 26$, $a=5$ and $\xi = 2$, the simulated CME acquired the speed of 1160 km/s at 50 $R_\odot$. We insert the CME in the background on 12-July-2012 21:30 UT, which makes the height of the CME at this time to be 36 $R_\odot$. Using this setup, the simulated CME arrives 3 hrs after the observed CME. In a future study, we will try to find the best approach to treat this speed-matching problem using multiple events, so that the arrival time difference can be minimized as much as possible.

\begin{figure}[!htb]
\center
\includegraphics[scale=0.1,angle=0,width=9cm,keepaspectratio]{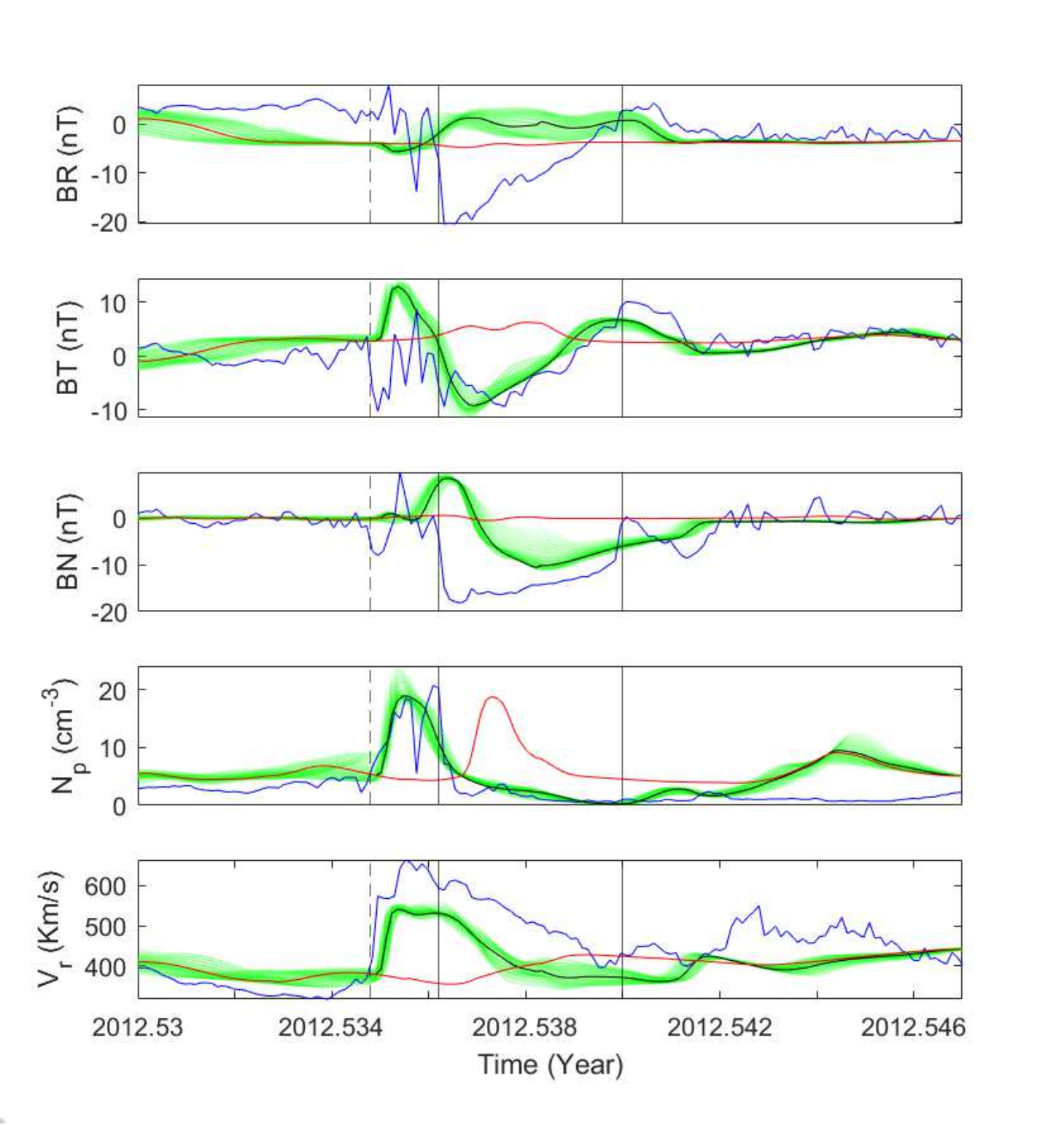}
\caption{Comparison of the 1 hr averaged OMNI solar wind data (blue) and  simulation results probed at Earth (black). Shock arrival is marked by the vertical dotted lines and the solid lines bound the magnetic cloud. The simulated CME arrived 3 hrs after the observed one. We also show the data probed at $\pm$ 5 degrees in latitude and longitude around Earth (green lines), to get an estimate of changes due to the possible error in the initial direction of the CME. Red lines show the solar wind solution when no CME was launched. We would like to emphasize that the $B_N$ values in RTN coordinates are approximately equal to the $B_z$ values in GSE coordinates}
\label{12July}
\end{figure}

In Fig. \ref{12July}, we show the comparison between the observations and simulations at Earth. We use the 1 hour averaged solar wind data provided by NASA/GSFC's OMNI data set through OMNIWeb \citep{King05}. The simulation results are probed along the  Earth trajectory. Also, to visualize the error that could be due to the error in the initial direction of the CME, we put multiple probes in $\pm$ 5 degrees latitude-longitude region and probe the plasma values there as well. They are shown by green lines in Fig. \ref{12July}. With red lines, we have plotted the simulation data when no CME was launched, to show the modulation of the solar wind by the CME. We note some differences between the model background solar wind and OMNI data ahead of the interplanetary shock arrival on 14 July 2012 (~2012.5348) that are possible sources of error in the sheath region behind the shock. First, the background magnetic field polarity is different between the model and observations, which could have contributed to the errors in the sheath magnetic fields. Secondly, the model solar wind is slightly denser and faster than observed (i.e., higher dynamic pressure). The propagation of CME is sensitive to the solar wind background \citep[e.g.][]{Temmer11,Lee13} , so it is important to improve the pre-CME solar wind reconstruction, possibly by further optimizing the WSA parameters. 

As for the comparison between CME values, we find that there is a good agreement in the density and $B_T$ values, and a fair agreement in $B_N$ and speed values. Our model did not capture the negative $B_R$ values exhibited by the actual CME. We also notice that the CME properties do not differ much in the $\pm$ 5 degrees region. This CME has previously been modeled by \citet{Shen14} and \citet{Scolini19} using a magnetised plasma blob model and a spheromak model, respectively. We notice that magnetic field values due to our modified spheromak model agrees much better with observations compared to above mentioned studies.  {For example, \citet{Shen14} were not able to match any component of magnetic field whereas \citet{Scolini19} were able to match only one component of magnetic field in one of their runs. However, the arrival time was missed by about 12 hrs.} We believe that constraining both poloidal and toroidal field can be a reason for our model's  better results. We have shown in our parametric study that changing these values can have a significant effect on 1 AU properties of CMEs. Another advantage our model has is that we introduce only a part of spheromak in the solar wind, such that it represents a bent tube geometry. \citet{Farrugia95} have shown that a tube geometry works much better than the full spheromak geometry to reproduce magnetic cloud signatures at 1 AU. 

While our proposed model and its data constraining technique shows a significant improvement on existing models, this model still does not capture the magnetic field signature of CMEs perfectly, for example, the negative $B_R$ values are not reproduced by this model at the Earth in this case. Several factors, including errors in initial direction, tilt, magnetic fluxes, etc. can be responsible for this, and we plan a detailed study to understand the impact of errors in initial parameters. We also plan to do a detailed study of this model using multiple events to understand its applicability as a operational model for CME predictions. We believe that the application of a flux rope-based model like ours will make a major advantage on existing non-flux-rope models.

\section{Conclusions}\label{conclusion}
In this study, we demonstrated the application of our modified spheromak model to simulate CMEs in the inner heliosphere driven by the WSA coronal model providing time-dependent boundary conditions. We described a robust technique to introduce the flux rope into the domain. We presented a parametric study to see the effect of model parameters on the  {simulated} CME evolution. They are summarized again below:
\begin{itemize}
    \item  {Initially, fast CMEs are decelerated and slow CMEs are accelerated. This has been confirmed by the observations as well.}
    \item  {In our simulations, the CME speed increases with the poloidal flux but does not depend on the toroidal flux. We have shown that this is because the pressure gradient and magnetic tension forces in a bent flux rope geometry, such as ours, are much more dependent on the poloidal flux than on the toroidal flux. Since actual CMEs have similar geometry, we may expect them to show similar behavior.}
    \item The energy multiplier $\xi$ can be used to control the CME speed, since the speed depends on $\xi$ linearly.  {However, $\xi$ should be kept less than 3 to avoid unnatural thermal pressure in the initial flux rope.};
    \item The simulated CME's speed increases with the decrease in the initial size of the flux rope.  {This is another parameter constraining the speed of simulated CMEs}
\end{itemize}

A CME-CME collision scenario, common during solar maxima can also be simulated using this model. We showed this in Sec. \ref{sec_coll}. Finally, in Sec. \ref{sec_12July}, we simulate the 12 July 2012 CME and compare the 1 AU properties with the observations. We find that our model was able to match the arrival time and plasma parameters with reasonable accuracy. It takes about 30 minutes of computation to simulate a CME propagation to 1 AU, on a $150\times128\times256$ resolution using 1200 CPUs. With this robustness, this model can be used as an operational tool for space weather prediction.

The authors acknowledge support from NASA grant 80NSSC19K0008. TS acknowledges the graduate student support from NASA Earth and Space Science Fellowship. TS and NVP acknowledge the support from the UAH IIDR grant 733033. We also acknowledge NSF PRAC award OAC-1811176 and related computer resources from the Blue Waters sustained-petascale computing project. Supercomputer allocations were also provided on SGI Pleiades by NASA High-End Computing Program award SMD-16-7570 and on Stampede2 by NSF XSEDE project MCA07S033. We acknowledge the NASA/GSFC Space Physics Data Facility's OMNIWeb for the solar wind and IMF data used in this study. We also used SOHO and STEREO coronagraph data from https://lasco-www.nrl.navy.mil and stereo-ssc.nascom.nasa.gov respectively. SDO EUV and magnetogram data has been taken from http://jsoc.stanford.edu/ajax/exportdata.html.

\bibliography{Singh2019}

\end{document}